\pdfoutput=1
\documentclass[11pt,twoside,a4paper,cmspaper,final,collab]{cms-tdr}
\def\svnVersion{6605a46}\def\svnDate{2025/11/17}\def\cmsCernNoTag{CERN-EP-2025-128}\def\cmsCernDate{\today}\def\cmsMessage{Published in the Journal of High Energy Physics as \href{http://dx.doi.org/10.1007/JHEP11(2025)038}{\doi{10.1007/JHEP11(2025)038}.}}
\begin{document}\cmsNoteHeader{JME-23-001}

\newcommand{\PS}{{\HepParticle{S}{}{}}\xspace}
\newcommand{\PY}{{\HepParticle{Y}{}{}}\xspace}
\renewcommand{\PT}{{\HepParticle{T}{}{}}\xspace}
\newcommand{\PTpr}{{\HepParticle{\PT}{}{\prime}}\xspace}
\newcommand{\PR}{{\HepParticle{R}{}{}}\xspace}
\providecommand{\PWst}{{\HepParticle{\PW}{}{\ast}}\xspace}
\newcommand{\mSD}{\ensuremath{m_\mathrm{SD}}}
\newcommand{\cmsTable}[1]{\resizebox{\textwidth}{!}{#1}}
\newcommand{\WKK}{\ensuremath{\PW_\mathrm{KK}}\xspace}

\cmsNoteHeader{JME-23-001}
\title{A method for correcting the substructure of multiprong jets using the Lund jet plane}

\date{\today}

\abstract{

Many analyses at the CERN LHC exploit the substructure of jets to identify heavy resonances produced with high momenta that decay into multiple quarks and/or gluons. This paper presents a new technique for correcting the substructure of simulated large-radius jets from multiprong decays. The technique is based on reclustering the jet constituents into several subjets such that each subjet represents a single prong, and separately correcting the radiation pattern in the Lund jet plane of each subjet using a correction derived from data. The data presented here correspond to an integrated luminosity of 138\fbinv collected by the CMS experiment between 2016--2018 at a center-of-mass energy of 13\TeV. The correction procedure improves the agreement between data and simulation for several different substructure observables of multiprong jets. This technique establishes, for the first time, a robust calibration for the substructure of jets with four or more prongs, enabling future measurements and searches for new phenomena containing these signatures. 
}
\hypersetup{
pdfauthor={CMS Collaboration},
pdftitle={A method for correcting the substructure of multiprong jets using the Lund jet plane},%
pdfsubject={CMS},
pdfkeywords={CMS, jets, substructure, tagging, Lund}}

\maketitle

\section{Introduction}

Many analyses of proton-proton ($\Pp\Pp$) collision data from the CERN LHC rely crucially on distinguishing hadronic decays of massive particles with a large Lorentz boost reconstructed as a single large-radius (large-$R$) jet from the overwhelming background of jets produced by quantum chromodynamics (QCD).
Large-$R$ jets are typically reconstructed using the infrared- and collinear-safe anti-\kt algorithm~\cite{Cacciari:2008gp, Cacciari:2011ma} with a large value of the distance parameter, $R \geq 0.8$.
Large-$R$ jets originating from heavy resonances will typically contain multiple `prongs', localized clusters of energy within the collection of final state particles of the jet, corresponding to the fragmentation and hadronization of the multiple colored partons from the resonance decay.
In contrast, jets from QCD are characterised by a hard core surrounded by soft and collinear radiation, and thus contain only a single prong.
Observables based on the substructure of the jet, i.e., the distribution of the energy among the jet's constituents, are used to reject QCD background and select events enhanced in signal-like characteristics~\cite{Thaler:2010tr,Komiske:2017aww,softdrop}.
Modern methods make extensive use of machine learning to achieve optimal discrimination performance \cite{CMS_tagging,CMS_ParticleNet,ATLAS_tagging,ATLAS_top_tagging}.
These methods are generally developed and optimized using Monte Carlo (MC) simulations, which, despite continual advances, are known to have deficiencies in the modeling of substructure observables~\cite{ATLAS_softdrop_meas,CMS_softdrop_meas}.
This mismodeling causes several difficulties for physics analyses making use of jet substructure. 

Analyses searching for a signal that produces a multiprong jet often use selection criteria based on jet substructure observables, referred to as jet tagging, to help remove the large QCD background.
However, it can be difficult to estimate the efficiency of such a selection in the signal process because of the known mismodeling in simulation.
Typically, a sample of representative standard model (SM) jets with similar substructure to the signal process is used for measuring the efficiency of the selection in data and for computing a scale factor (SF) to correct the efficiency in simulation~\cite{CMS_tagging,BTV-16-002, ATLAS_higgs_tagging,ATLAS_tagging}.
However, only two-prong jets, due to $\PW$ and $\PZ$ bosons, and three-prong jets, due to top ($\PQt$) quarks, are produced in sufficient abundance in the SM, meaning this procedure cannot be performed for signals producing jets with four or more prongs (high-pronged). 
For such processes, including beyond-the-SM (BSM) particles producing large-$R$ jets with a large number of prongs~\cite{Bai:2018jsr, Aguilar-Saavedra:2017iso, Agashe:2016rle, Agashe:2016kfr} and decays of boosted Higgs bosons via $\PW\PWst$ or $\PZ\PZst$ to four quarks, there is no suitable SM process to use as a calibration tool.
This lack of a suitable proxy poses a significant challenge in interpreting searches for such signatures.

This paper describes a new approach to correct the simulation of large-$R$ multiprong jets and solve these challenges.
To derive the calibration and validate its efficacy, data are used corresponding to an integrated luminosity of 138\fbinv collected by the CMS experiment in 2016--2018 at a center-of-mass energy of 13\TeV.
The proposed technique is generic, applicable to jets containing a large number of prongs, provided that each prong of the jet is produced by an SM quark. 
Our technique is based on reclustering the jet into a fixed number of subjets, such that each subjet encapsulates the showering of a single quark from the initial hard interaction. 
The radiation patterns in each of these individual prongs can be corrected separately and then combined to obtain a correction for the full jet. 
The procedure relies on the assumption that the radiation pattern of individual prongs in low- and high-prong jets are similar, such that the modeling of high-prong jets can be corrected from a per-prong calibration derived from low-prong jets. 
We validate this assumption in studies using data and simulation.

The correction for a single prong is based on an estimate of the Lund jet plane (LJP) density~\cite{Dreyer:2018nbf} of subjets in data and simulation.
The LJP provides a two-dimensional representation of the radiation pattern inside jets.
It encodes the phase space density of different types of splittings, in terms of momentum transfer and angular separation, inside the radiation pattern of a jet.
The ratio between data and simulation of the LJP density is used to reweight each subjet based on the likelihood of its splittings.
Finally, the correction for the full radiation pattern of the multiprong jet is given by the product of corrections for each of its subjets.
In this paper, only prongs originating from quarks are considered.
In principle, this method can be applied to gluon-initiated subjets, but a different correction must be derived since gluon-initiated subjets are expected to have a different radiation pattern from those originating from quarks.
Multi-gluon jets would have larger color reconnection effects between the different prongs, which complicates the factorization of the large-$R$ correction into separate per-prong corrections. 

Because this method corrects the underlying radiation pattern of a simulated multiprong jet, it can also be used to obtain a more accurate estimate of the efficiency of substructure-based selection criteria, as well as the corresponding uncertainty.
The correction is quantified as an SF, defined as the ratio of the efficiency as computed from a more accurate method (\eg, after the LJP substructure correction or measured using an SM proxy) to the efficiency in simulation.
The SFs are used to compare the LJP method with existing methods using abundant SM proxy processes, and good agreement is observed.
Although the LJP method generally has larger uncertainties than existing methods for these low-pronged signals, it can be additionally applied to high-pronged signals which are not covered by existing methods.

The per-splitting correction is derived from a sample of large-$R$ jets originating from boosted \PW bosons 
(referred to as \PW jets) in a region enriched in leptons+jets \ttbar events, i.e, where the \PW boson from one top decays leptonically and the \PW boson from the other top decays hadronically.
Boosted \PW bosons are used to derive the correction because their decays produce a clean sample of quark-initiated subjets, whereas most samples of single-prong jets are a mixture of quark- and gluon-initiated jets, which would complicate the extraction of the correction. 
Once the ratio of LJP densities is derived, it is applied to the same sample of \PW jets, and it is shown to simultaneously improve  the modeling of multiple substructure observables.
The method is further validated in jets originating from boosted top quarks, showing that a correction derived from two-prong jets can be applied to jets with more than two prongs. 

Since the efficacy of the method on jets with four or more prongs cannot be directly checked in data, additional studies are performed in simulation to validate its usage on such jets.
Simulated samples of both lepton+jets \ttbar and BSM particles producing jets with four to six prongs are generated using two different showering algorithms, \PYTHIA~\cite{Sjostrand:2014zea} and \HERWIG~\cite{Webber:1983if,Gieseke:2003rz}.
The lepton+jets \ttbar samples are used to define a \PYTHIA-to-\HERWIG correction, using an procedure analogous to that used to correct simulation to data.
These corrected \PYTHIA samples are compared with the native \HERWIG samples, and the two agree within the uncertainties of the method, which validates the applicability of the method to processes with higher prong multiplicities.

Differential measurements of the LJP density for quark and gluon jets have been performed for $R = 0.4$ jets by the ATLAS Collaboration~\cite{ATLAS_Lund} and for $R = 0.4$ and $R = 0.8$ jets by the CMS Collaboration~\cite{CMS_Lund}.
Measurements of the LJP density for jets originating from the hadronic decays of boosted top quarks and \PW bosons have also been performed by the ATLAS Collaboration~\cite{ATLAS:LundWtop}.
This work is the first time information from the LJP has been used to perform a correction to simulated jets. 
A search by the CMS Collaboration for resonances decaying to three \PW bosons~\cite{triboson} developed a calibration procedure for the case of four-pronged jets  
using top quark jets with a hard gluon emission as an SM proxy for the decay of a boosted massive BSM scalar, the radion, to two \PW bosons.
However, the method required extensive validation to demonstrate the applicability of the SM proxy object to the signal process, and it does not generalize easily to jets with higher prong multiplicities. 
The generality of the LJP method allows for improved calibration of similar four-pronged jets, as well enabling for the first time the calibration of jets with even higher prong multiplicities. 
This new calibration capability will allow for the proper interpretation of searches for signatures containing jets with high prong multiplicities.
It has been employed for the calibration of multiprong jets in a prior search by the CMS Collaboration for dijet resonances with anomalous substructure \cite{CMS:2024nsz}, which set exclusion limits on a variety of signal models producing jets from two to six prongs. 

The paper is structured as follows.
The CMS detector and event reconstruction procedures are described in Section \ref{sec:CMS_reco}.
Section \ref{sec:samples} describes the data and simulation samples used to derive and validate the method.
Section \ref{sec:event_selection} describes the selection criteria applied to select the  $\ttbar$ lepton+jets sample, and Section \ref{sec:derive_correction} details how the LJP density ratio is extracted from this sample. 
Section \ref{sec:apply_correction} describes the application of the correction to simulations of \PW and \PQt jets, and validations in data demonstrating its efficacy.
Section \ref{sec:herwig_study} details studies performed in simulation which validate the applicability of the method to jets with more than three prongs.
Uncertainties are discussed in Section \ref{sec:uncs}.
Comparisons of calibrated tagging efficiencies from the LJP method and others are given in Section \ref{sec:eff_examples}, and a summary is given in Section \ref{sec:summary}.

\section{The CMS detector and event reconstruction }
\label{sec:CMS_reco}

The central feature of the CMS apparatus is a superconducting solenoid of 6\unit{m} internal diameter, providing a magnetic field of 3.8\unit{T}.
Within the solenoid volume are a silicon pixel and strip tracker, a lead tungstate crystal electromagnetic calorimeter (ECAL), and a brass and scintillator hadron calorimeter (HCAL), each composed of a barrel and two endcap sections.
Forward calorimeters extend the pseudorapidity ($\eta$) coverage provided by the barrel and endcap detectors.
Muons are measured in gas-ionization detectors embedded in the steel flux-return yoke outside the solenoid.
A more detailed description of the CMS detector, together with a definition of the coordinate system used and the relevant kinematic variables, is reported in Refs.~\cite{CMS:2008xjf,CMS:2023gfb}.

The silicon tracker used in 2016 measured charged particles within the pseudorapidity range $\abs{\eta} < 2.5$.
For nonisolated particles of transverse momentum (\pt) of $1 < \pt < 10\GeV$ and $\abs{\eta} < 1.4$, the track resolutions were typically 1.5\% in \pt and 25--90 (45--150)\mum in the transverse (longitudinal) impact parameter~\cite{CMS:2014pgm}.
At the start of 2017, a new pixel detector was installed~\cite{Phase1Pixel} that measured particles up to $\abs{\eta} < 3.0$ with typical resolutions of 1.5\% in \pt and 20--75\mum in the transverse impact parameter~\cite{DP-2020-049} for nonisolated particles of $1 < \pt < 10\GeV$.
Muons are measured in the pseudorapidity range $\abs{\eta} < 2.4$, with detection planes made using three technologies: drift tubes, cathode strip chambers, and resistive-plate chambers.
The single-muon trigger efficiency exceeds 90\% over the full $\eta$ range, and the efficiency to reconstruct and identify muons is greater than 96\%.
Matching muons to tracks measured in the silicon tracker results in a relative transverse momentum resolution, for muons with \pt up to 100\GeV, of 1\% in the barrel and 3\% in the endcaps.
The \pt resolution in the barrel is better than 7\% for muons with \pt up to 1\TeV~\cite{CMS:2018rym}.

Events of interest are selected using a two-tiered trigger system.
The first level, composed of custom hardware processors, uses information from the calorimeters and muon detectors to select events at a rate of around 100\unit{kHz} within a fixed latency of about 4\mus~\cite{CMS:2020cmk}.
The second level, known as the high-level trigger (HLT), consists of a farm of processors running a version of the full event reconstruction software optimized for fast processing, and reduces the event rate to around 1\unit{kHz} before data storage~\cite{CMS:2016ngn}.

The global event reconstruction, also called particle-flow (PF) event reconstruction~\cite{CMS:2017yfk}, aims to reconstruct and identify each individual particle in an event, with an optimized combination of all subdetector information.
In this process, the identification of the particle type (photon, electron, muon, charged hadron, neutral hadron) plays an important role in the determination of the particle direction and energy.
Photons are identified as ECAL energy clusters not linked to the extrapolation of any charged particle trajectory to the ECAL.
Electrons are identified as a primary charged particle track and potentially many ECAL energy clusters corresponding to this track extrapolation to the ECAL and to possible bremsstrahlung photons emitted along the way through the tracker material.
Muons are identified as tracks in the central tracker consistent with either a track or several hits in the muon system, and associated with calorimeter deposits compatible with the muon hypothesis.
Charged hadrons are identified as charged-particle tracks neither identified as electrons, nor as muons.
Finally, neutral hadrons are identified as HCAL energy clusters not linked to any charged-hadron trajectory, or as a combined ECAL and HCAL energy excess with respect to the expected charged-hadron energy deposit.

The energy of photons is obtained from the ECAL measurement.
The energy of electrons is determined from a combination of the track momentum at the main interaction vertex, the corresponding ECAL cluster energy, and the energy sum of all bremsstrahlung photons attached to the track.
The energy of muons is obtained from the corresponding track momentum.
The energy of charged hadrons is determined from a combination of the track momentum and the corresponding ECAL and HCAL energies, corrected for the response function of the calorimeters to hadronic showers.
Finally, the energy of neutral hadrons is obtained from the corresponding corrected ECAL and HCAL energies.

For each event, hadronic jets are clustered from these reconstructed particles using the infrared- and collinear-safe anti-\kt algorithm~\cite{Cacciari:2008gp, Cacciari:2011ma}.
Jets reconstructed with a distance parameter of $R=0.8$ (AK8 jets) are used for reconstructing heavy resonance decays, whereas jets with $R=0.4$ (AK4 jets) are used to reconstruct jets originating from single quarks or gluons. 
Jet momentum is determined as the vectorial sum of all particle momenta in the jet, and is found from simulation to be, on average, within 5 to 10\% of the true momentum over the entire \pt spectrum and detector acceptance. 
Additional $\Pp\Pp$ interactions within the same or nearby bunch crossings (pileup) can contribute additional tracks and calorimetric energy depositions to the jet momentum. 
The pileup-per-particle identification algorithm (PUPPI)~\cite{Bertolini:2014bba, Sirunyan:2020foa} is used for AK8 jets to mitigate the effect of pileup at the reconstructed-particle level, making use of local shape information, event pileup properties, and tracking information.
A local shape variable is defined, which distinguishes between collinear and soft diffuse distributions of other particles surrounding the particle under consideration.
The former is attributed to particles originating from the hard scatter and the latter to particles originating from pileup interactions.
Charged particles identified to be originating from pileup vertices are discarded. For
each neutral particle, a local shape variable is computed using the surrounding charged particles compatible with the primary vertex within the tracker acceptance ($\abs{\eta} < 2.5$), and using both charged and neutral particles in the region outside of the tracker coverage.
The momenta of the neutral particles are then rescaled according to their probability to originate from the primary interaction vertex deduced from the local shape variable, superseding the need for jet-based pileup corrections~\cite{Sirunyan:2020foa}.

Jet energy corrections are derived from simulation to bring the measured response of jets to that of particle level jets on average.
In situ measurements of the momentum balance in dijet, $\text{photon}+\text{jet}$, $\PZ+\text{jet}$, and multi-jet events are used to determine any residual differences between the jet energy scale in data and in simulation~\cite{CMS:2016lmd}. 
Additional selection criteria are applied to each jet to remove jets potentially dominated by instrumental effects or reconstruction failures.
The jet energy resolution amounts typically to 15--20\% at 30\GeV, 10\% at 100\GeV, and 5\% at 1\TeV~\cite{CMS:2016lmd}.

The missing transverse momentum vector, \ptvecmiss, is computed as the negative vector sum of the transverse momenta of all the PF candidates in an event, and its magnitude is denoted as \ptmiss~\cite{CMS:2019ctu}.
The \ptvecmiss is modified to account for corrections to the energy scale of the reconstructed jets in the event.
Anomalous high-\ptmiss events can be due to a variety of reconstruction failures, detector malfunctions, or noncollision backgrounds.
Such events are rejected by event filters that are designed to identify more than 85--90\% of the spurious high-\ptmiss events with a mistagging rate less than 0.1\%~\cite{CMS:2019ctu}.

\subsection{Reconstruction of the Lund jet plane}

To identify the subjets inside the AK8 jets, an additional reclustering procedure is performed.
For a specified number of subjets, corresponding to the expected number of prongs inside the jet, the constituents of the AK8 jets are reclustered into the chosen number of subjets using the exclusive-\kt algorithm~\cite{Ellis:1993tq}. 
Unlike the standard \kt clustering, which uses the distance parameter $R$ for its stopping criteria, the exclusive version terminates only when the specified number of subjets is reached.
The energy of these subjets is corrected with the same jet energy correction factors that are used for the corresponding AK8 jet.
Once the subjets have been determined, the clustering tree within each subjet's \kt clustering history is used to obtain the splittings used to build the LJP density.
The correction is performed on the primary LJP, so only splittings along the hardest branch of the clustering history are used.

Studies in simulation are performed to assess the subjet identification performance of the \kt algorithm as compared to the Cambridge--Aachen (CA)~\cite{Dokshitzer:1997in, Wobisch:1998wt}  or anti-\kt algorithms.
Subjets from the \kt algorithm are found to more closely align with the prongs from the hard process as compared to subjets from the CA or anti-\kt algorithms.
This is consistent with the use of exclusive-\kt algorithm subjets in other substructure methods, such as $N$-subjetiness~\cite{Thaler:2010tr}.
Reference~\cite{Dreyer:2018nbf} suggests building the LJP based on splittings from the CA algorithm and cautions that using splittings from the \kt algorithm may produce unwanted features.
A modified procedure, in which the subjets from the \kt algorithm are further reclustered with the CA algorithm to obtain the clustering tree for the LJP, shows no significant performance improvements, and therefore it is not used in these studies.

\section{Data and MC samples}
\label{sec:samples}

The data sample used in the analysis is based on $\Pp\Pp$ collisions at $\sqrt{s} =13\TeV$ collected from 2016 to 2018, corresponding to an integrated luminosity of 138\fbinv.
The integrated luminosities for the 2016, 2017, and 2018 data-taking years have 1.2--2.5\% individual uncertainties~\cite{CMS-LUM-17-003,CMS-PAS-LUM-17-004,CMS-PAS-LUM-18-002}, whereas the overall uncertainty for the 2016--2018 period is 1.6\%.

Simulated $\Pp\Pp$ collision events are generated at $\sqrt{s} = 13 \TeV$ using various generators.
Except where noted, the parton shower and hadronization is simulated using \PYTHIA~8.240 with the CP5 tune~\cite{CMS:2019csb}.
The NNPDF3.1 parton distribution functions (PDFs) at next-to-next-to-leading order (NNLO) accuracy~\cite{NNPDF:2017mvq} are used for all samples.
The detector response is simulated using \GEANTfour~\cite{GEANT4:2002zbu}. 

The \ttbar process is generated with the next-to-leading order (NLO) generator \POWHEG~v2~\cite{Nason:2004rx,Frixione:2007nw,Alioli:2010xd}.
An alternative \ttbar sample, used only for validation studies, is created based on the same \POWHEG configuration but instead interfaced to \HERWIG v7.2.2 for showering.
The lepton+jets \ttbar sample is divided into three components based on the angular distance, defined as $\Delta R = \sqrt{\smash[b](\Delta \eta)^2 + (\Delta \phi)^2}$, where $\phi$ is the azimuthal angle in radians, between the generator-level decay products of the hadronically decaying top quark and the momentum vector of the reconstructed AK8 jet. 
Events in which the two quarks from the $\PW$ boson decay and the $\PQb$ quark all have $\Delta R<0.8$ from the AK8 jet  are considered `$\PQt$-matched'.
Events in which the two quarks from the $\PW$ boson decay are within $\Delta R < 0.8$ of the reconstructed AK8 jet, but the $\PQb$ quark is not, are considered `$\PW$-matched'.
All other events are considered `unmatched'.

The $\PQt\PW$ process is simulated at NLO accuracy with \POWHEG, where the \PQt quark decays via the lepton+jet channel. 
Similarly to the \ttbar sample, events are categorized as `\PW-matched' if the two quarks from the \PW decay are within $\Delta R<0.8$ of the reconstructed AK8 jet, and `unmatched' otherwise. 

Simulated events of $\PW+\text{jets}$ and $s$-channel single \PQt quark production are generated using \MGvATNLO~2.6.5~\cite{Frederix:2012ps} 
at leading order (LO) accuracy, with the MLM jet-merging prescription~\cite{Alwall:2007fs}.
The single \PQt quark process in the $t$-channel is generated at NLO accuracy with \POWHEG. 
Events composed uniquely of jets produced through the strong interactions, referred to as QCD multi-jet events, 
and the $\PW\PW$, $\PW\PZ$, and $\PZ\PZ$ processes are generated using \PYTHIA at LO accuracy.  

The most accurate available cross section calculations are used to normalize the SM simulated samples, in most cases corresponding to NNLO accuracy in QCD~\cite{Czakon:2011xx, Campbell:2020fhf,PDF4LHCWorkingGroup:2022cjn, Gehrmann:2014fva,Cascioli:2014yka,Campbell:2011bn}.
The \pt spectrum of top quarks in \ttbar events is reweighted to account for effects due to missing higher-order corrections in MC simulation~\cite{CMS:2016oae}.

Several BSM processes are generated to evaluate the LJP correction procedure on jets with large prong multiplicity. 
All BSM samples are simulated using \MGvATNLO at LO accuracy, and two versions of each sample are produced, using \PYTHIA or \HERWIG for the parton shower and hadronization.
Two-pronged jets are simulated using a sample of $\PX\to \PS \PY$, where \PX is a vector boson with $m(\PX)=3\TeV$, \PY is a vector boson with $m(\PY)=170\GeV$, and $\PS$ is a scalar with $m(\PS)=170\GeV$.
Both $\PY$ and $\PS$ decay to a pair of SM quarks. 
A sample of vector-like top quarks, $\PTpr$, is used to study five-pronged jets. 
The $\PTpr$ quarks have a mass of $m(\PTpr)=400\GeV$, decay via $\PTpr\to \PZ\PQt$, and are pair-produced via a heavy vector boson with a mass of 5\TeV~\cite{Okada:2012gy, Buchkremer:2013bha}. 
Six-pronged jets are simulated using a sample of heavy scalars, \PH, with mass $m(\PH)=400\GeV$, which decay via $\PH\to \PQt\PAQt$ and are pair produced via a heavy Randall--Sundrum graviton with a mass of 3\TeV~\cite{Carvalho:2014lsg}. 
Finally, a sample containing both two- and four-pronged jets is produced using a Kaluza--Klein excitation of the \PW boson, \WKK, decaying to a \PW boson and a radion, R, which itself decays to two \PW bosons, with $m(\WKK)=2.5\TeV$ and $m(\PR)=200\GeV$~\cite{Agashe:2016rle,Agashe:2017wss}.

\section{Selection of lepton+jets \texorpdfstring{\ttbar}{ttbar} events}
\label{sec:event_selection}

A data sample of large-$R$, multiprong jets with well-defined composition is needed in order to derive the LJP density ratio corrections and to test their performance. 
For this purpose, a sample of lepton+jets \ttbar events is used, targeting \PW jets.
A similar sample was previously used by the CMS Collaboration for calibration of the performance of large-$R$ jet taggers~\cite{CMS_tagging}.
The event selection seeks to pick out events where one of the top quarks decays into a muon, a neutrino, and a \PQb jet, resulting in large \ptmiss.
The other top quark decays hadronically, producing a large-$R$ jet on the other side of the event that is used to derive the LJP density correction.

Events are selected requiring the presence of a muon, $\ptmiss>50\GeV$, an AK4 jet, and an AK8 jet.
The muon is required to have $\pt > 60\GeV$, $\abs{\eta} < 2.4$, and an unsigned transverse impact parameter, $\abs{d_{xy}}$, less than 2\unit{mm}.
It is also required to pass identification criteria based on the number of hits observed in the tracker, the response of the muon detectors, and a set of matching criteria between muon track parameters, as measured by the inner tracker and muon detectors~\cite{CMS:2018rym}.
To suppress nonprompt muons coming from heavy flavor decays, the muon must be isolated from other particles in a cone of size $\Delta R = 0.3$.
More details on the muon identification and reconstruction is described in Refs.~\cite{CMS:2018rym,CMS_highpt_muons}.

The AK4 jet is required to have $\pt > 25\GeV$ and $\abs{\eta} < 2.4$, and must be identified as a \PQb jet.
The \PQb tagging is performed with the \textsc{DeepJet} algorithm, which combines information from the PF candidates and secondary vertices ~\cite{BTV-16-002, Bols:2020bkb, CMS-DP-2018-058}.
The \PQb tagging working point used corresponds to an average efficiency of $\sim$80\% for selecting genuine \PQb jets, and a $\sim$1\%  (2\%) rate of misidentifying light flavored or gluon (charm) jets.
The AK4 jet is also required to have a $\phi$ angular separation with respect to the selected muon of less than 2.
To remove overlap with the muon, it is required that the AK4 jet be $\Delta R > 0.4$ away from the selected muon.

The AK8 jet is required to have $\pt > 225 \GeV$ and $\abs{\eta} < 2.4$. 
The AK8 jet is also required to have a $\phi$ angular separation with respect to the selected muon of more than 2. 

\begin{figure}[!th]
  \centering
  \includegraphics[width = 0.6\textwidth]{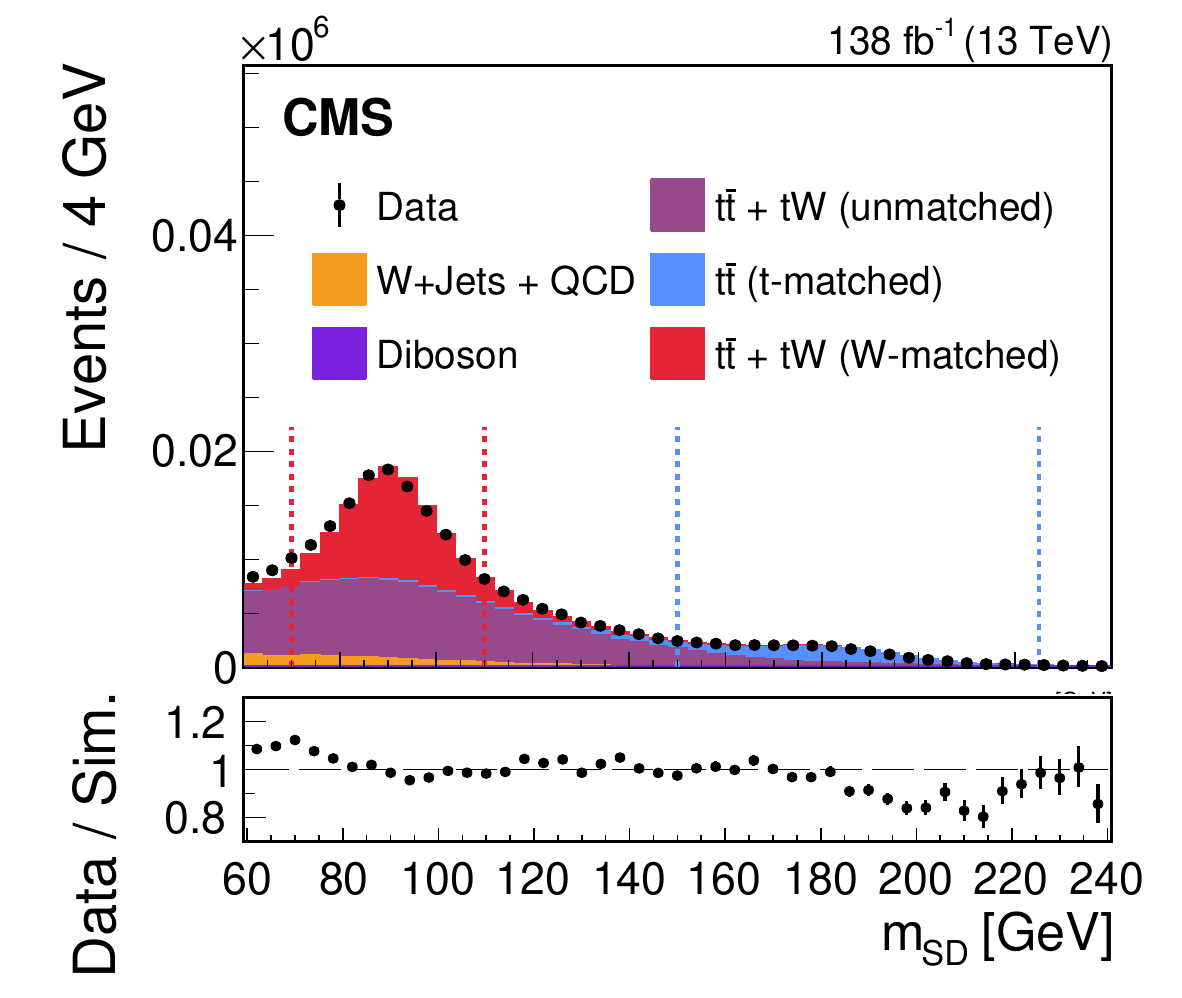}
  \caption{ 
    The distribution of the soft-drop mass for AK8 jets in the lepton+jets \ttbar region prior to the LJP density correction.
    The number of simulated events has been scaled to match the observed number of data events. 
    The lower panel shows the ratio between the observed data and the simulated estimates.
    Only statistical uncertainties are shown as vertical bars on the data points.
    The red (blue) dashed vertical lines denote the mass range of 70--110\GeV (150--225\GeV), which defines the \PW (\PQt) region used in the analysis.
}
  \label{fig:full_msoftdrop}
\end{figure}

The soft-drop (SD) algorithm~\cite{softdrop}, a generalization of the modified mass drop algorithm~\cite{Butterworth:2008iy,Dasgupta:2013ihk},
with the angular exponent $\beta = 0$ and soft threshold $z = 0.1$, is applied to the AK8 jet to reconstruct the jet mass (\mSD).
The SD algorithm removes soft and collinear radiation from the jet to improve the mass resolution for jets originating from heavy resonances.
A comparison of the $\mSD$ distributions between data and simulation, for AK8 jets in the lepton+jets \ttbar region, is shown in Fig. \ref{fig:full_msoftdrop}.

Events are split into two regions based on the soft-drop mass of the AK8 jet. 
The \PW region targets \PW jets, with soft-drop mass between 70 and 110\GeV, whereas the \PQt region targets top quark jets, with soft-drop mass between 150 and 225\GeV. 
The \PQt region additionally requires the AK8 jet to have $\pt>500\GeV$. 
Only events from the \PW region are used to derive the LJP density data/simulation ratios that define the correction, detailed in Section~\ref{sec:derive_correction}, whereas the \PQt region is used to validate the performance of the method.

\section{Definition of LJP density ratios and their extraction from data}
\label{sec:derive_correction}

The first step of the correction procedure is to recluster the large-$R$ jet into a specified number of subjets, chosen such that each subjet encapsulates a single prong.
The radiation of each prong can then be corrected separately, allowing one to build up a correction for the full jet from each of the subjets. 
The correction for each subjet is based on the ratio of LJP densities of subjets in data to that of simulation. 
This ratio is derived from the sample of \PW jets in the \PW region. 
This region is chosen because it consists of a relatively pure sample of large-$R$ jets of a known number of prongs (two)  and has a sufficient event count.

The reclustering proceeds via the steps specified in Section~\ref{sec:CMS_reco} to obtain the subjets and their corresponding splittings.
Once the splittings are obtained, they are used to construct the LJP densities for both data and simulation. 
In each splitting, the lower (higher) \pt branch is denoted as the emission (emitter).
The kinematics of a splitting is described by the splitting angle of the branching ($\Delta$) and transverse momentum of the emission
relative to the  emitter (\kt).
All splittings in the primary LJP of each subjet are used in the correction.
The LJP density is typically binned in the logarithm of these two quantities, because the soft and collinear divergences of QCD produce emissions that are uniform in $\ln(\kt/\text{GeV})$ and $\ln(1/\Delta$).
The correction is also binned in the \pt of the subjet, i.e., separate LJP densities are defined for subjets of different \pt. 
Specifically, the LJP density is binned in $\ln(\kt/\text{GeV})$ and $\ln(0.8/\Delta$), with 19 and 14 bins covering the ranges $-1.0$ to $5.3$ and $-4.4$ to $6.9$, respectively.
Six subjet \pt bins are used, with lower edges at 10, 65, 110, 175, 240, and 300\GeV, where the last bin includes all subjets with $\pt>300\GeV$. 

\begin{figure}[!th]
  \centering
  \includegraphics[width = 0.48\textwidth]{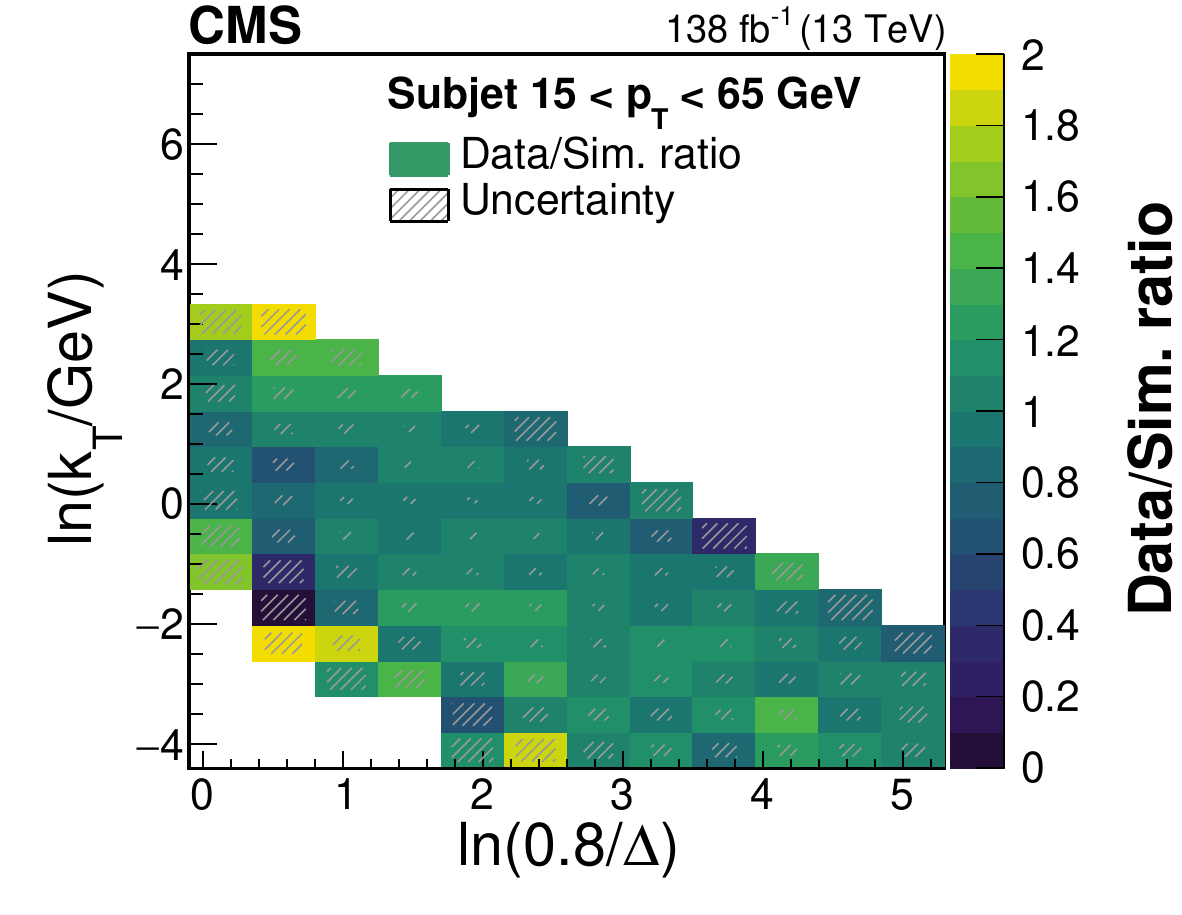}
  \includegraphics[width = 0.48\textwidth]{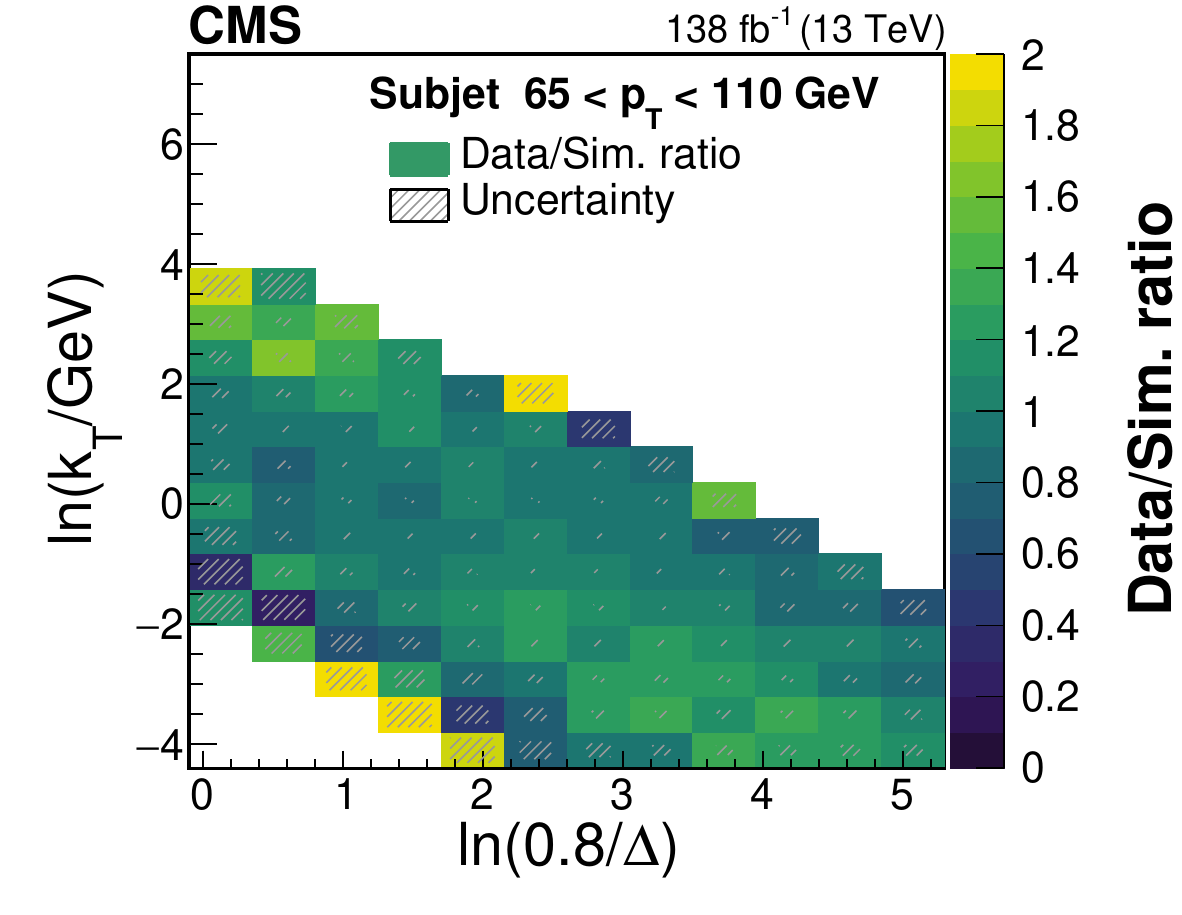}\\
  \includegraphics[width = 0.48\textwidth]{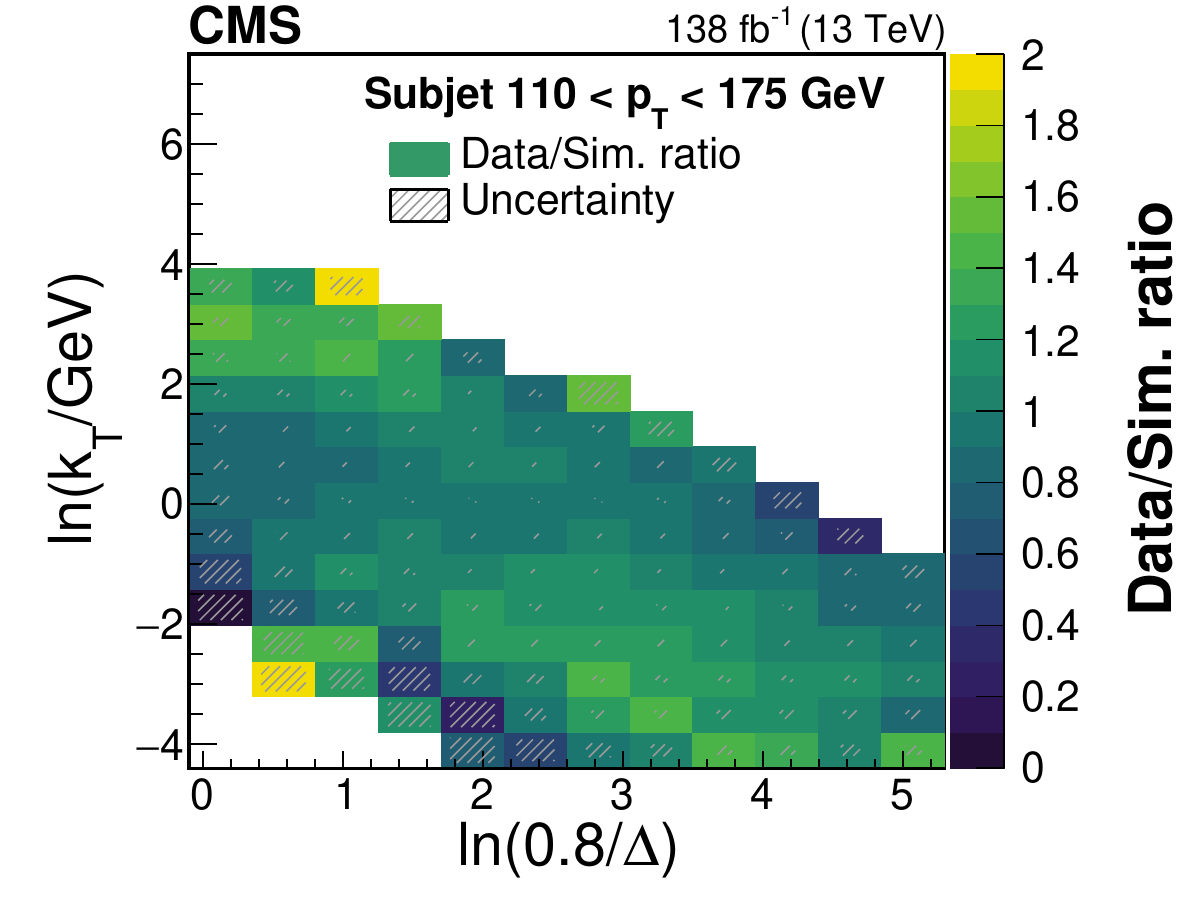}
  \includegraphics[width = 0.48\textwidth]{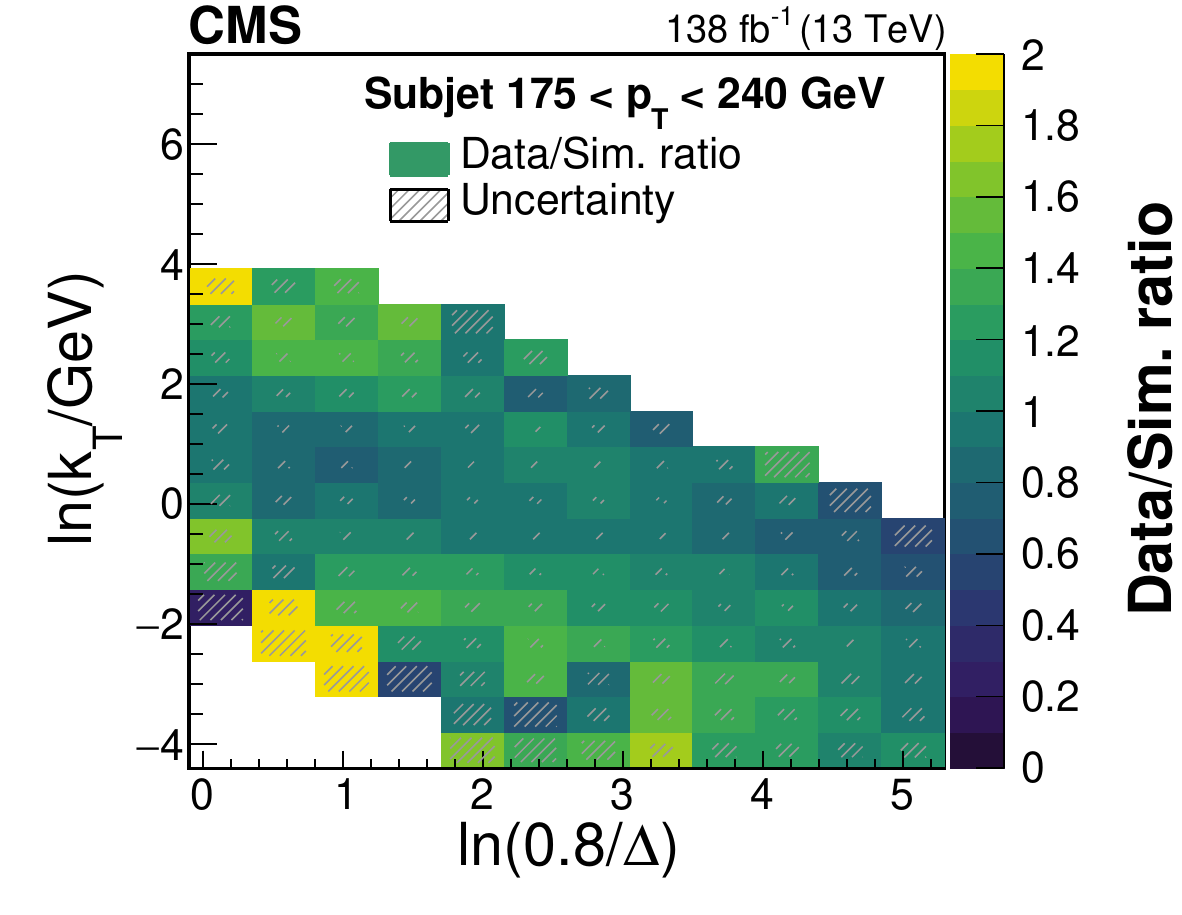}\\
  \includegraphics[width = 0.48\textwidth]{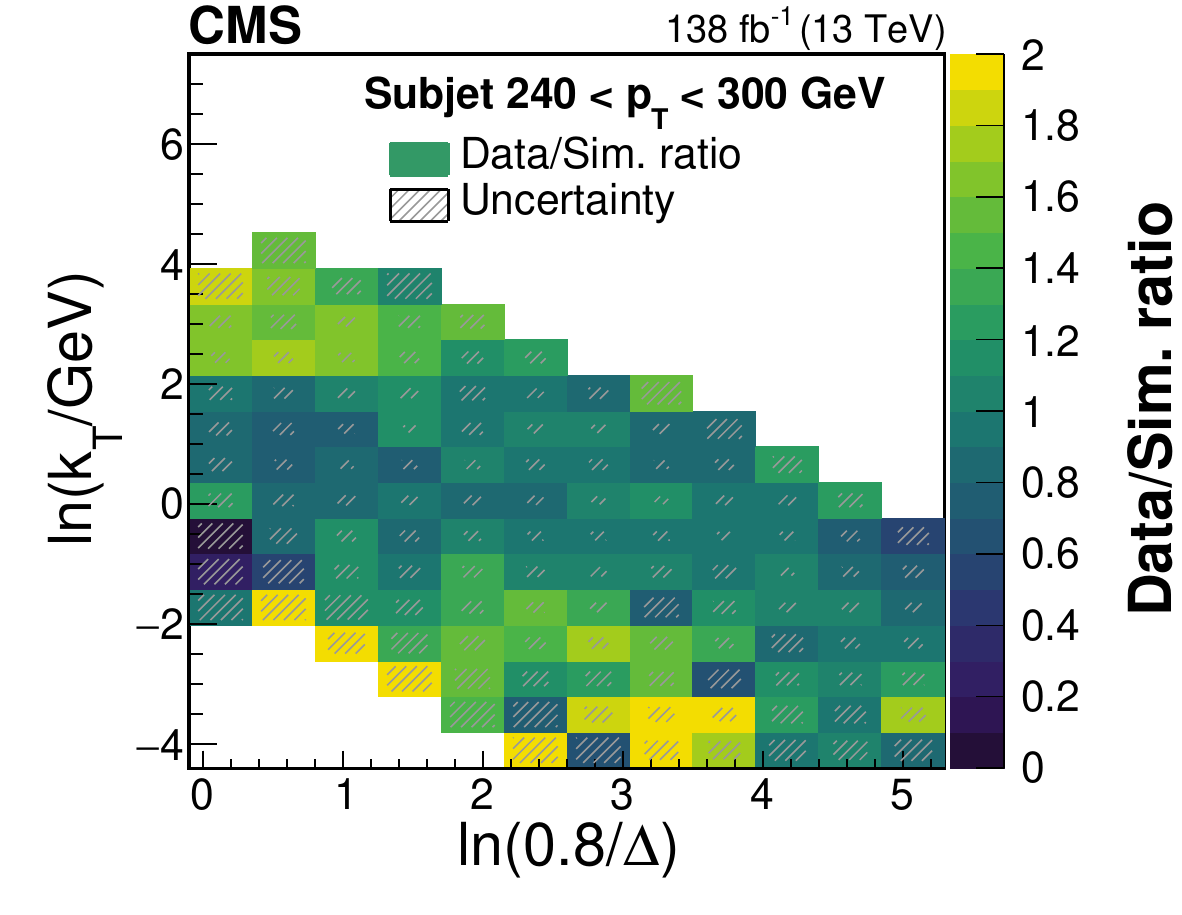}
  \includegraphics[width = 0.48\textwidth]{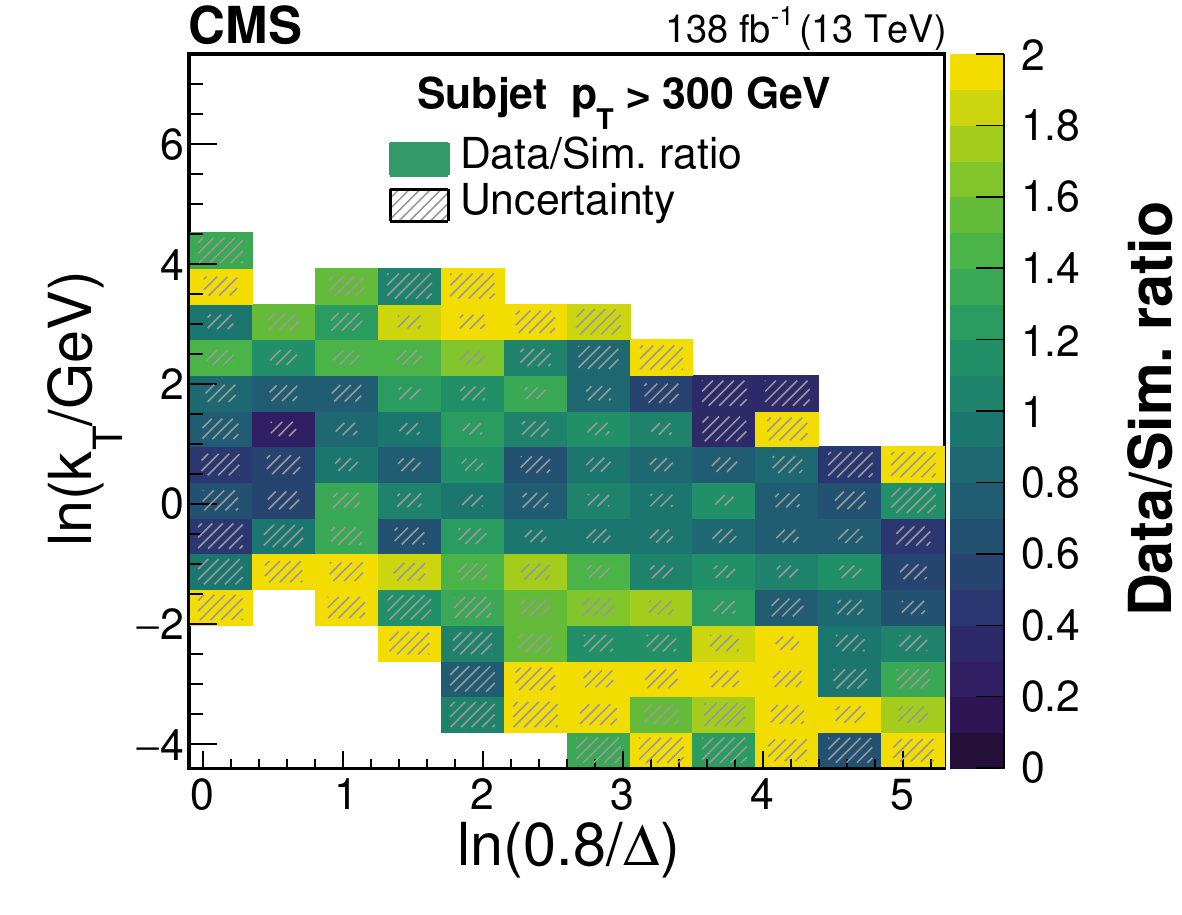}
  \caption{ 
    Ratios of the LJP densities between data and simulation in the six subjet \pt bins. 
    Bins with no data or simulation events are shown as white; in the application of the correction, they are assigned a ratio value of unity and an uncertainty of 100\%.
    The ratio values have been restricted to an upper limit of 2 for visualization purposes.
    The combined statistical and systematic uncertainty in the ratio is represented by the area of the hatched region in each bin.
    The fractional size of the hatched region in each bin represents the uncertainty in the measured ratio value in that bin,
    \eg, for bins in which the hatched region covers half of the area, the fractional uncertainty in the measured ratio is 50\%. 
    A description of the considered systematic uncertainties is given in Section \ref{sec:uncs}.
    The ratios are used to build the corrections to the substructure of a subjet. 
  }

  \label{fig:LP_ratio}
\end{figure}

\begin{figure}[!th]
  \centering
  \includegraphics[width = 0.49\textwidth]{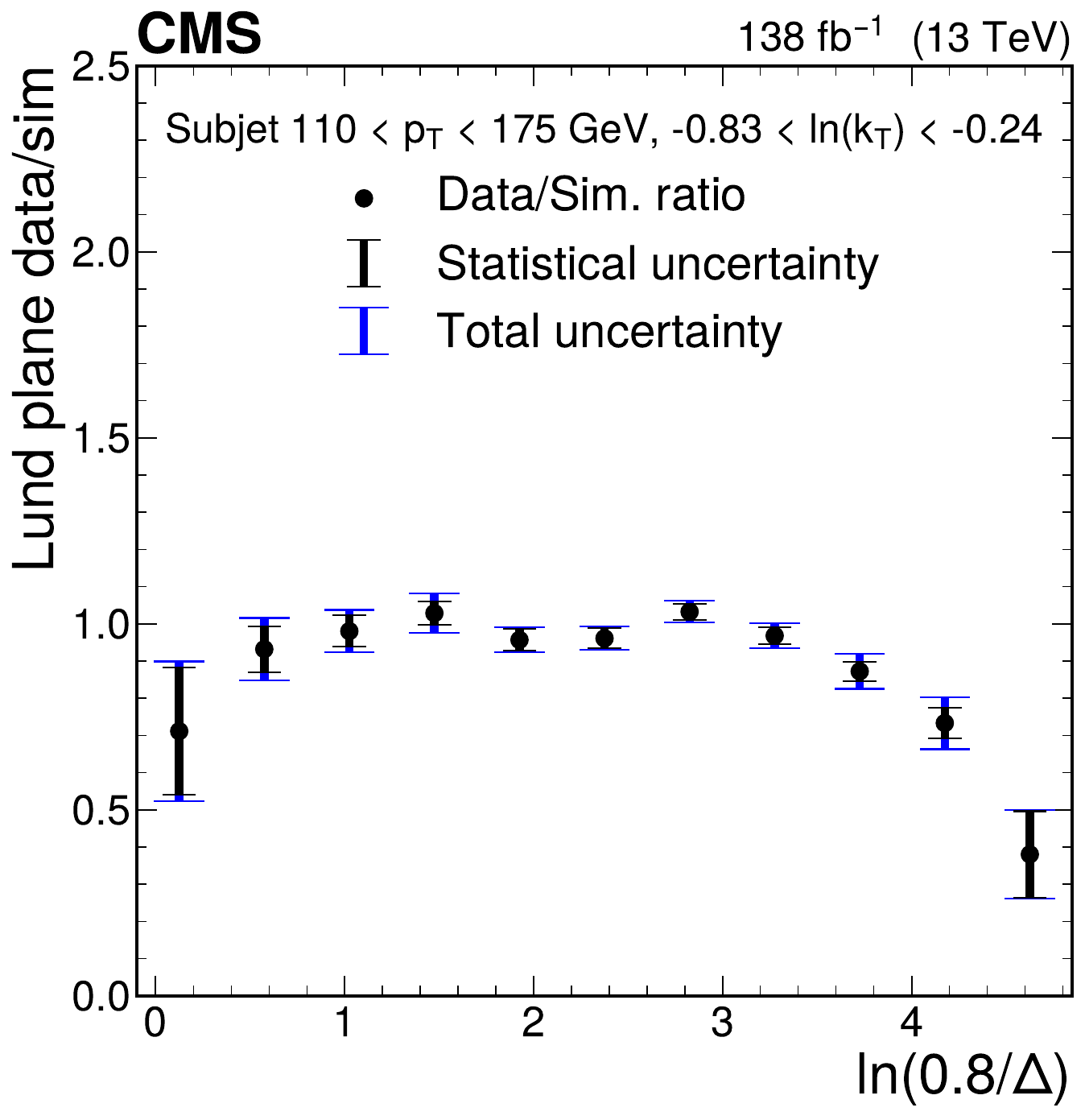}
  \includegraphics[width = 0.49\textwidth]{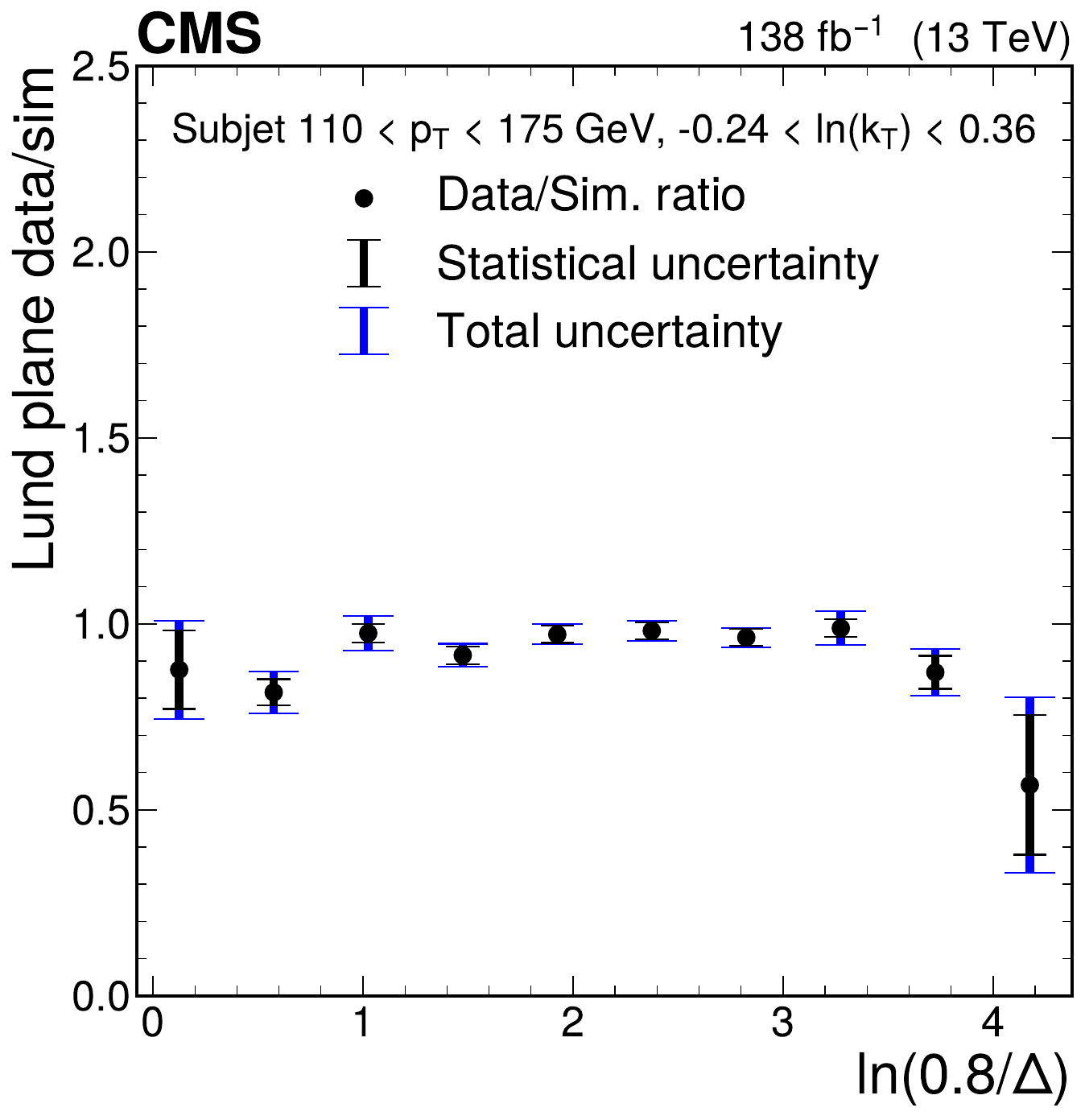}
  \includegraphics[width = 0.49\textwidth]{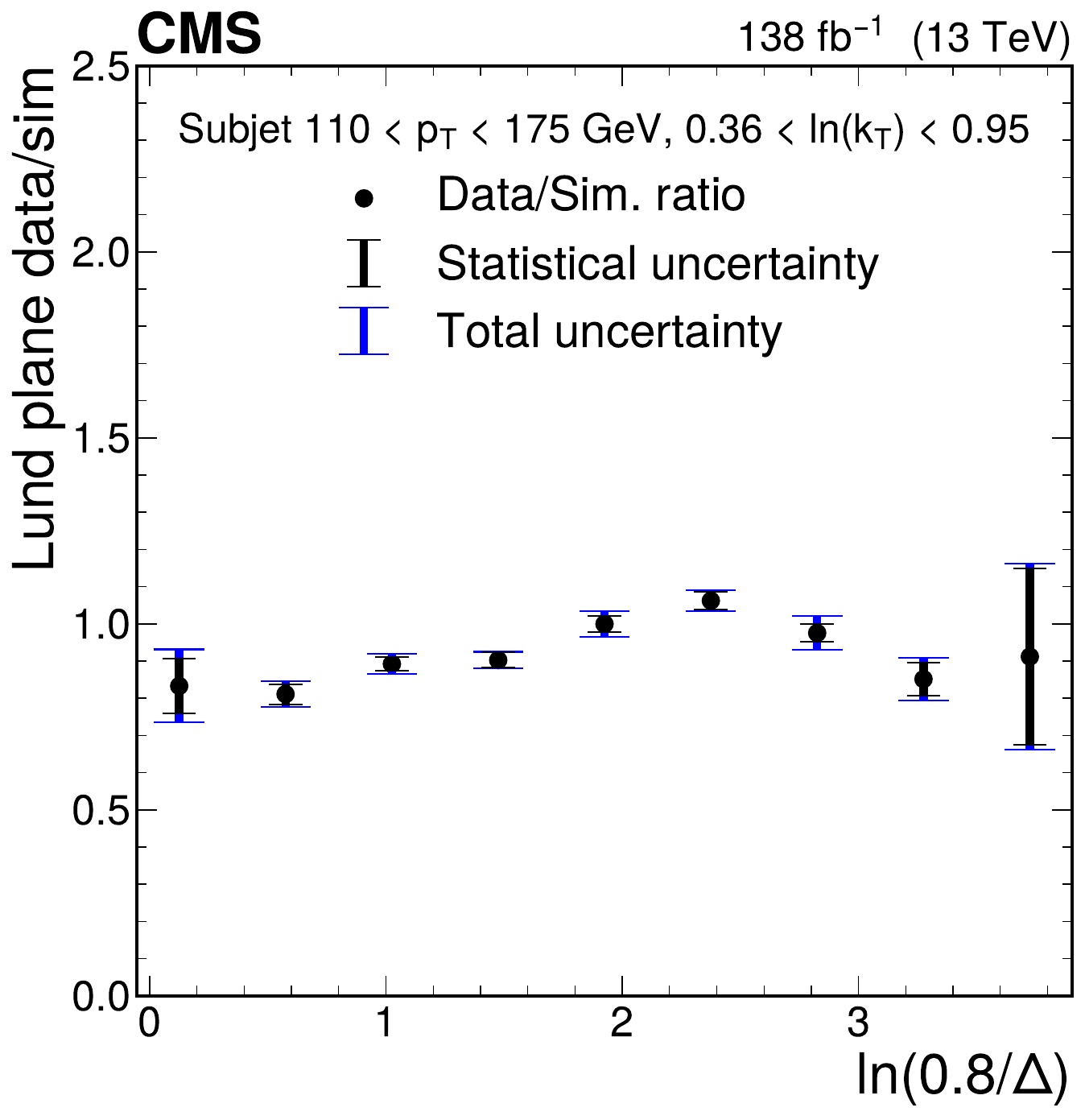}
  \includegraphics[width = 0.49\textwidth]{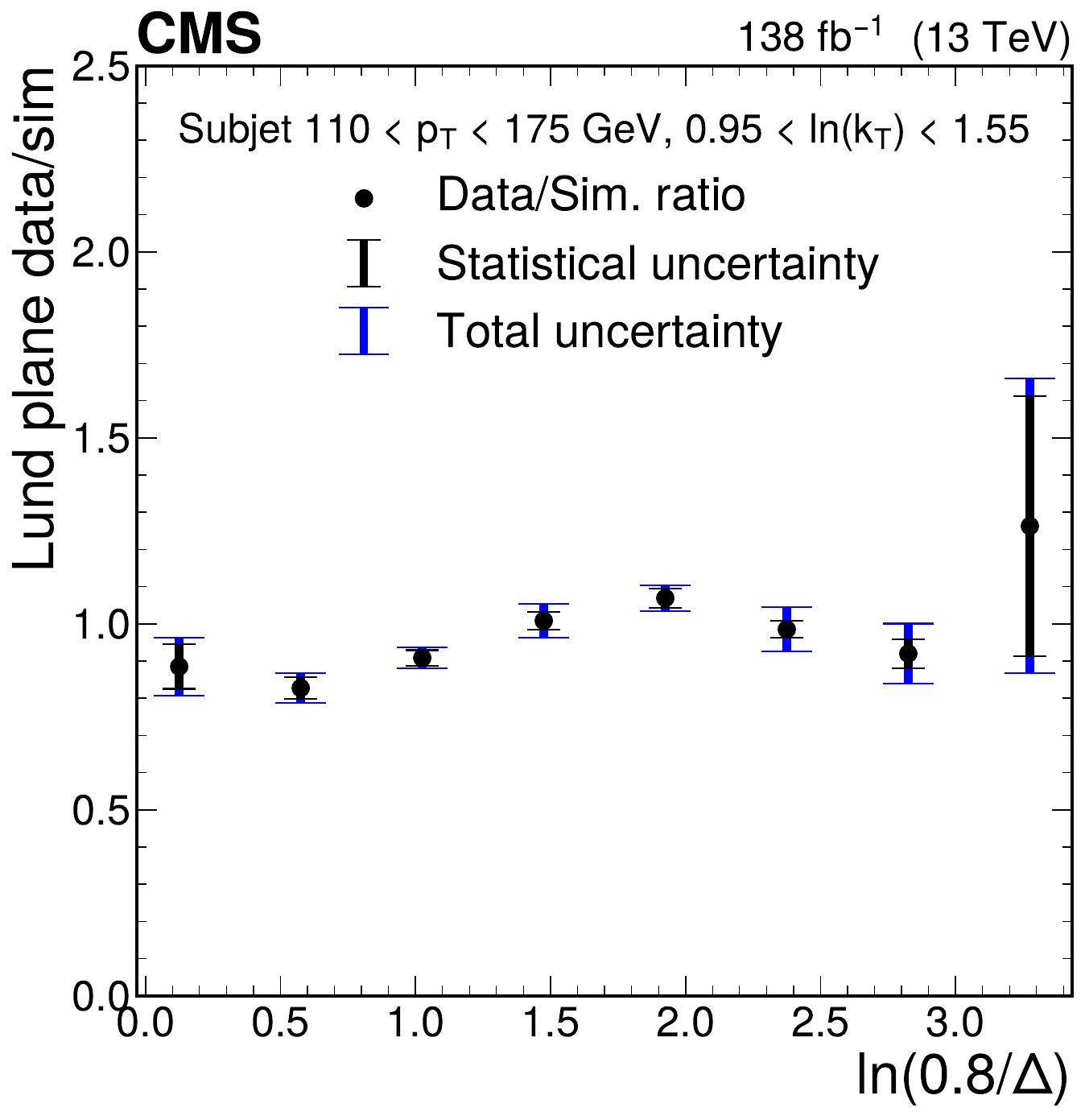}
  \caption{ 
    Ratios of the LJP densities between data and simulation projected into one dimension. 
    The ratio is shown as a function of $\ln(0.8/\Delta)$ for several \kt bins for the subjet \pt bin 110--175\GeV.
    Statistical uncertainties are shown as the black error bars, and the combined statistical and systematic uncertainties are shown as the 
    blue error bars.
    The statistical uncertainties dominate the uncertainty in most bins. 
  }

  \label{fig:LP_1D}
\end{figure}

For each \pt bin, three LJP densities are constructed from the \PW region in different samples: one for signal simulation (i.e., \PW-matched \ttbar and $\PQt\PW$ events), 
one for background simulation (all other processes), and one for data. 
The reclustering step is performed assuming 2-pronged jets. 
The LJP density of simulated background events is subtracted from the data LJP density to obtain an estimate of the LJP density of subjets within \PW jets. 
The simulation LJP density and data-minus-background LJP density are then normalized to the number of subjets in the sample,  so that they encode the frequency of splittings for a subjet. 
Finally, the ratio of the normalized data-minus-background LJP density and the signal simulation LJP density is computed.
The LJP density ratio and its uncertainty is shown in Fig.~\ref{fig:LP_ratio}, and several one-dimensional slices are shown in Fig.~\ref{fig:LP_1D}. 
Details of the systematic uncertainty are on the ratio are given in Section \ref{sec:uncs}.

To reduce the impact of outliers, each bin of the ratio is restricted to the range 0.2--5. 
There are few outliers outside of this range, all of which have large statistical uncertainties that are comparable in size to the central value. 
Statistical uncertainties from the limited number of data and simulation events are propagated to a statistical uncertainty in the final ratio.
Because of limited event counts and phase space restrictions, some bins within the two-dimensional plane have zero events in either simulation and data.
The ratio in these bins is taken to be unity with a 100\% uncertainty. 
As discussed in Section~\ref{sec:uncs}, these uncertainties are sub-dominant as compared to other uncertainties on the correction procedure.

\subsection{Extrapolation to higher \texorpdfstring{\pt}{pt}}
\label{subsec:pt_extrap}

The sample of \PW jets used only produces subjets in a limited \pt range. 
In order to apply the correction procedure to jets with subjet \pt higher than available in the \PW region, an extrapolation procedure is performed. 
The LJP densities are expected to have limited variation as a function of jet \pt. 
At leading order, the LJP density can be written as a function of $\overline{z} = \frac{\kt}{p_{T,\mathrm{subjet}} \Delta}$ \cite{Dreyer:2018nbf},
and in the limit of soft or wide-angle emissions, $\overline{z} \ll 1$, the density approaches a constant value.
The extrapolation to higher \pt is therefore defined as a power series in $\frac{1}{p_{T, \mathrm{subjet}}}$. 

For each bin in $\ln(\kt/\GeVns{})$ and $\ln(0.8/\Delta$), the correction factors in the six subjet \pt bins are included 
The procedure starts with a zeroth-order polynomial fit.
Higher-order polynomials are tested and adopted if they significantly improve the fit quality.
Specifically, the quality of the fit is quantified with a $\chi^2$ test, where the number of degrees of freedom is the number of subjet \pt bins with a nonzero entry for this particular $\ln(\kt/\GeVns{})$ and $\ln(0.8/\Delta)$ bin.
If the corresponding $p$-value is greater than 0.3 for a fit of a given order, the process terminates. 
If not, the polynomial order is increased by one, and a Fisher F-test~\cite{FisherTest} is performed to check if the higher-order fit is preferred at $p < 0.05$.
If the lower-order fit is preferred, the process terminates; otherwise, the process is repeated with the higher-order polynomial as the new baseline. 
This extrapolation is done independently for each $\ln(\kt/\GeVns{})$ and $\ln(0.8/\Delta)$ bin.
The procedure finds that a constant order is sufficient for nearly all ($\approx$97\%) bins, and a linear order is preferred by the rest.

The extrapolated values are used to reweight subjets with $\pt > 300\GeV$.
For each splitting in such a subjet, the corresponding $\ln(\kt/\GeVns{})$ and $\ln(0.8/\Delta)$ bin is determined, the extrapolation function describing the correction factor in that bin is evaluated at the given subjet \pt, and the resulting value is used as the correction factor.
Similar to the measured ratio values, the extrapolation correction factors are restricted to the range 0.2--5. 
The systematic uncertainties in the extrapolation are described in Section~\ref{subsec:pt_unc}.

\section{Correction application and validation on \texorpdfstring{\PW}{W} and top quark jets}
\label{sec:apply_correction}

Once the LJP density ratios are derived, they are applied to correct the substructure of simulated multiprong jet as follows:
\begin{enumerate}
  \item Using generator-level information from the simulation, determine how many quarks from the hard process decay are contained in the cone of the jet ($N$).
  \item Recluster the PF candidates of the jet into $N$ subjets using the exclusive \kt algorithm. Subjets matched to a generator-level quark within $\Delta R<0.2$ are used in the correction; the unmatched subjets are not used in the nominal correction, but are considered in the systematic uncertainties. 
  \item For all subjets matched to a generator-level quark, obtain the clustering history produced by the \kt algorithm. 
  \item For each splitting along the hardest branch of the clustering history with $\kt > 0.02 \GeV$, 
    look up the data/simulation correction factor based on the subjet's \pt and the splitting's \kt and $\Delta$.
  \item Multiply the correction factors for each splitting to obtain the total subjet correction.
  \item Multiply the correction factors for each matched subjet to obtain the total jet correction. The total jet correction is used as a reweighting factor, multiplying the event weight of the MC events. 
  \item If there are multiple jets in the event that are being corrected, multiply the weights of the individual jets to obtain the total event weight.
  \item \label{ite:norm} To preserve the overall normalization of the sample, divide the event weights by the mean reweighting factor across all events in the sample.
\end{enumerate}

\begin{figure}[!t]
  \centering
  \includegraphics[width = 0.99\textwidth]{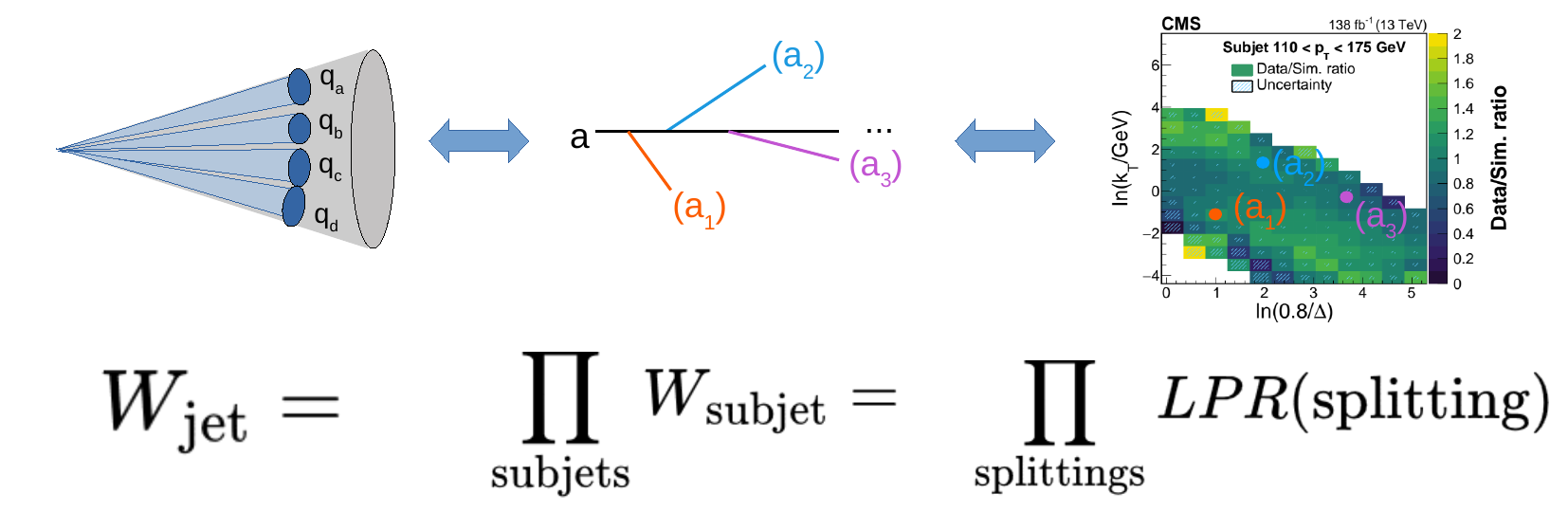}
  \caption{ 
    A graphical illustration of the correction procedure. First, the large-$R$ jet is reclustered into its subjets. 
    Then, the clustering history for each subjet is used to obtain an list of splittings from the primary LJP.
    For each splitting, the LJP density ratio is used as a correction factor. 
}
  \label{fig:flowchart}
\end{figure}

This procedure is illustrated graphically in Fig.~\ref{fig:flowchart}. 
The final jet weight is a product of many splitting-based reweighting factors, 
and therefore has an approximate log-normal distribution.
Due to properties of the log-normal distribution, even if the distribution of splitting weights has a mean of one, their product will not.
This necessitates the post-hoc correction of Step~\ref{ite:norm} to ensure the reweighting procedure preserves the sample normalization.

The $\kt > 0.02 \GeV$ requirement was determined based on the minimal momentum and spatial resolution of reconstructed charged hadrons, and tighter restrictions were found to reduce  the improvement in the data-simulation agreement of the correction. 
Variations of $\Delta R$ matching requirement between 0.15--0.25 were tried and produce minimal impact on the results of the correction. 

Because the correction is applied to the underlying radiation pattern of the jet, it is expected to improve the modeling of all substructure observables simultaneously.

\begin{figure}[!th]
  \centering
  \includegraphics[width = 0.49\textwidth]{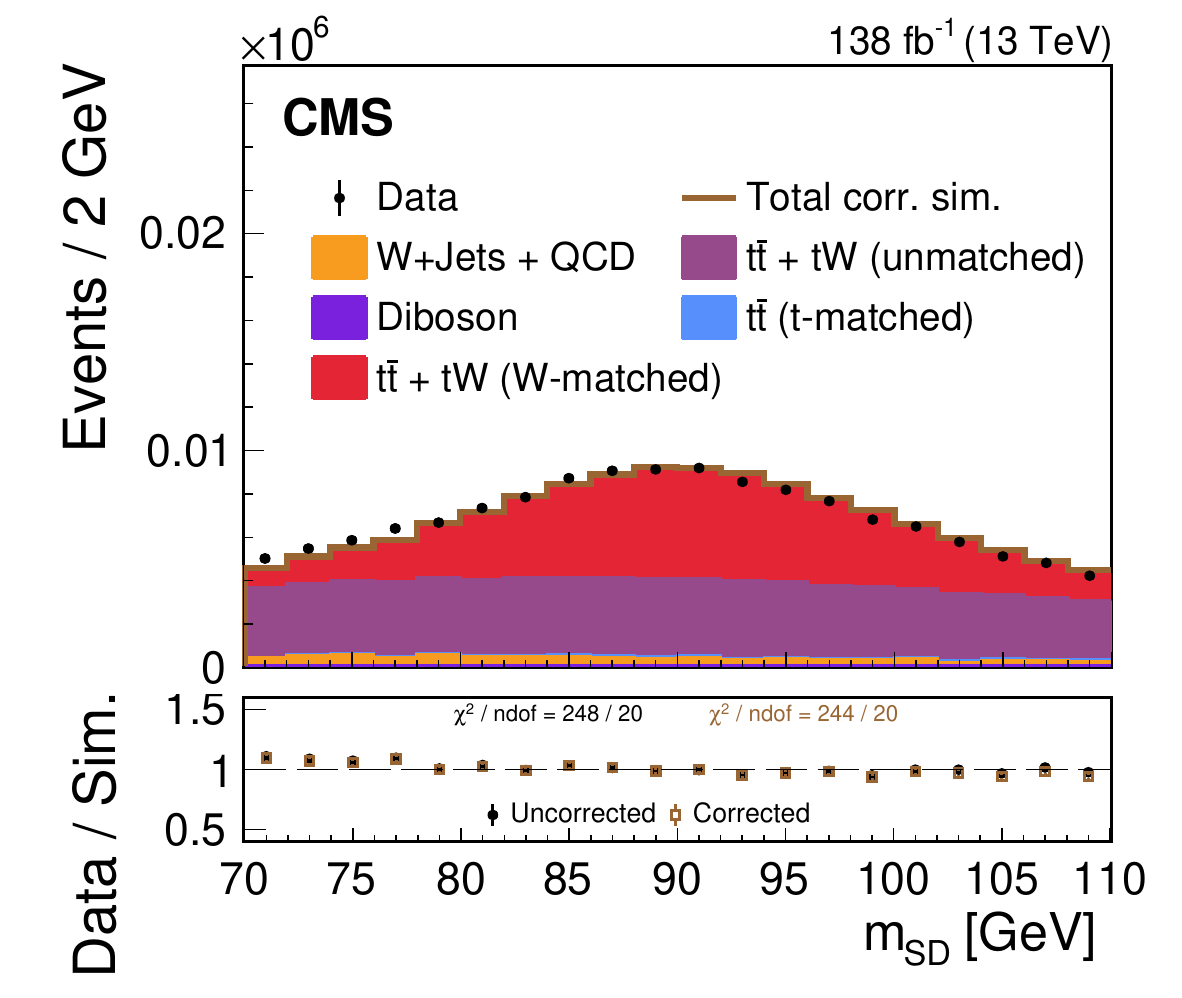}
  \includegraphics[width = 0.49\textwidth]{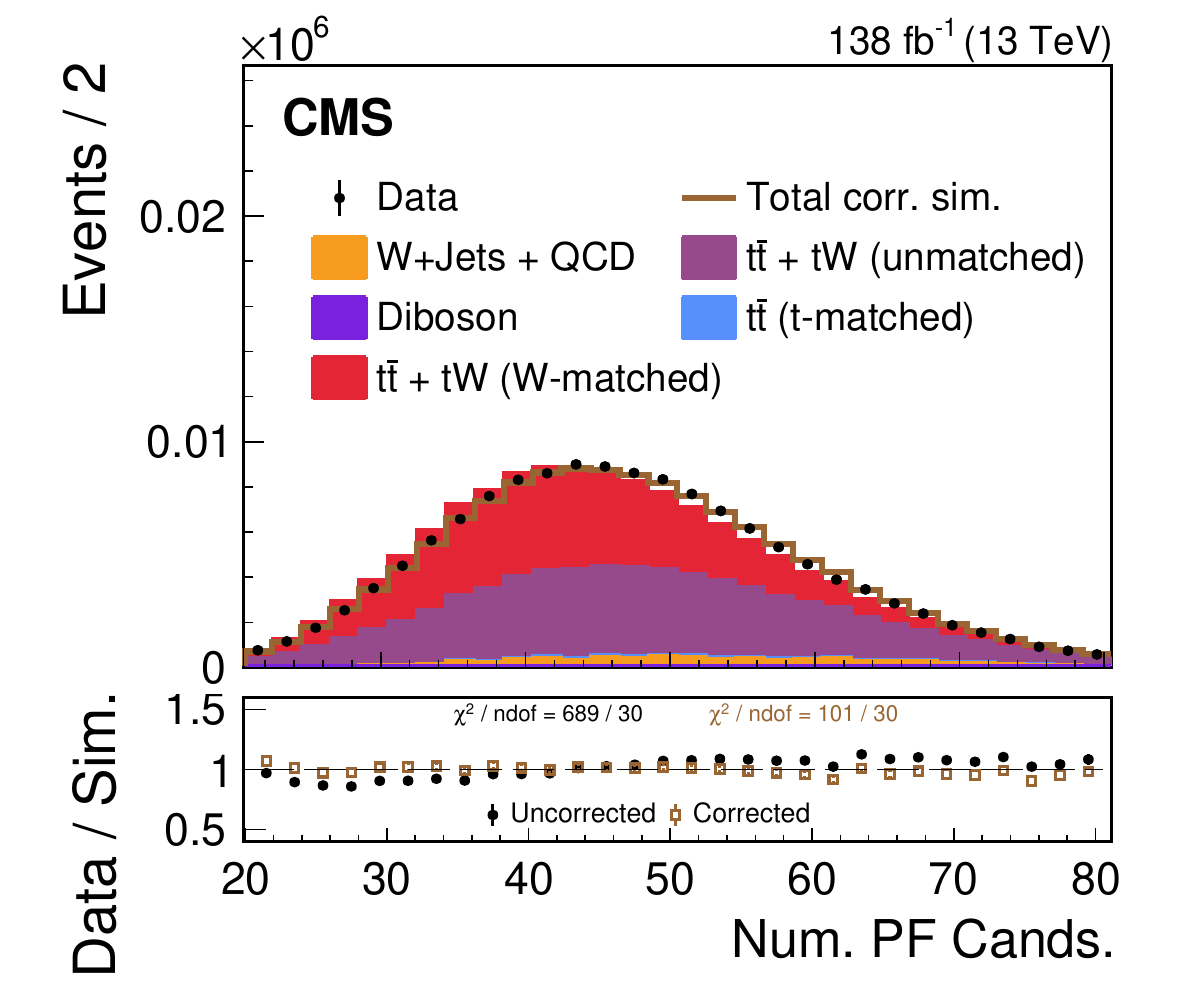} \\
  \includegraphics[width = 0.49\textwidth]{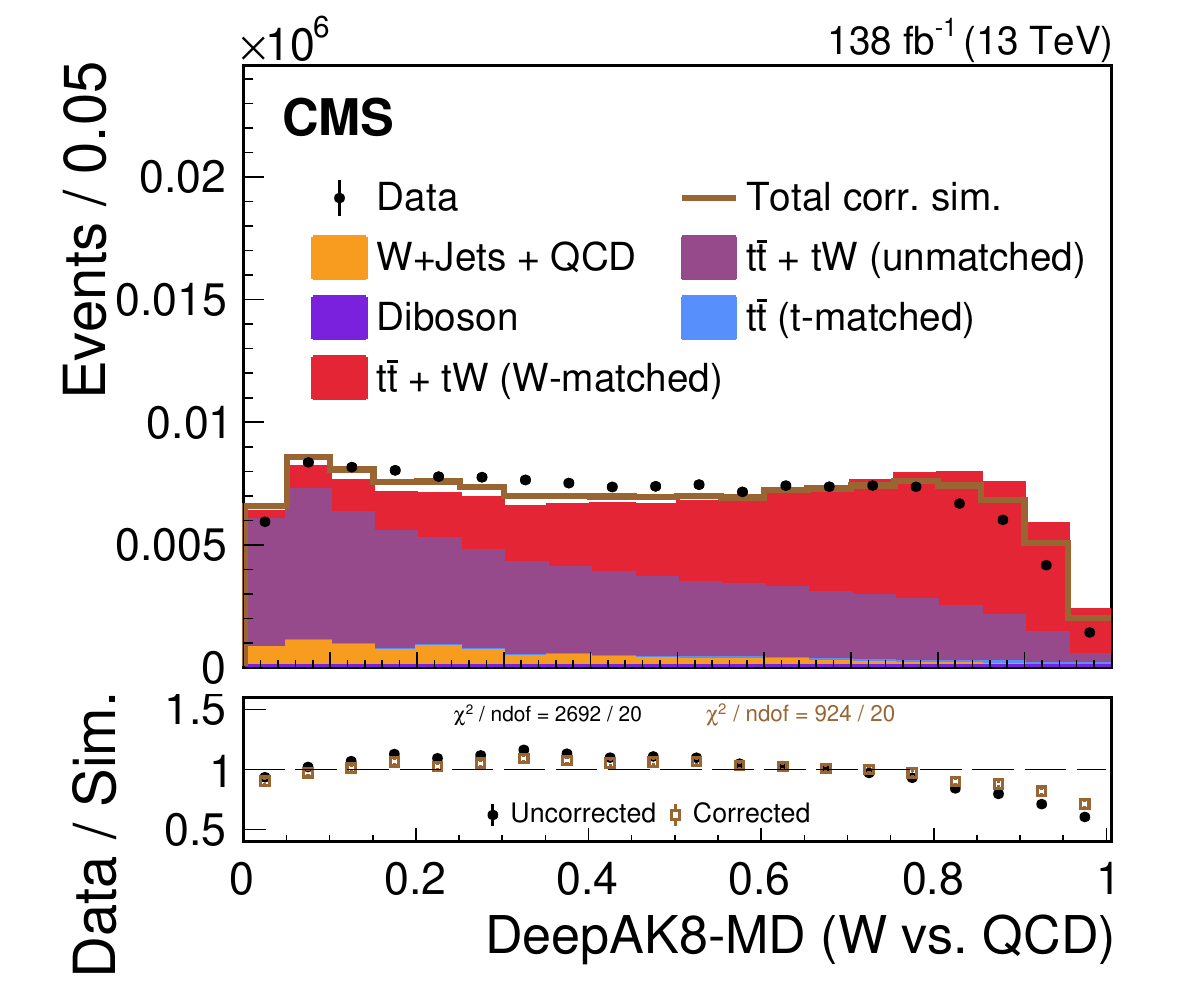} 
  \includegraphics[width = 0.49\textwidth]{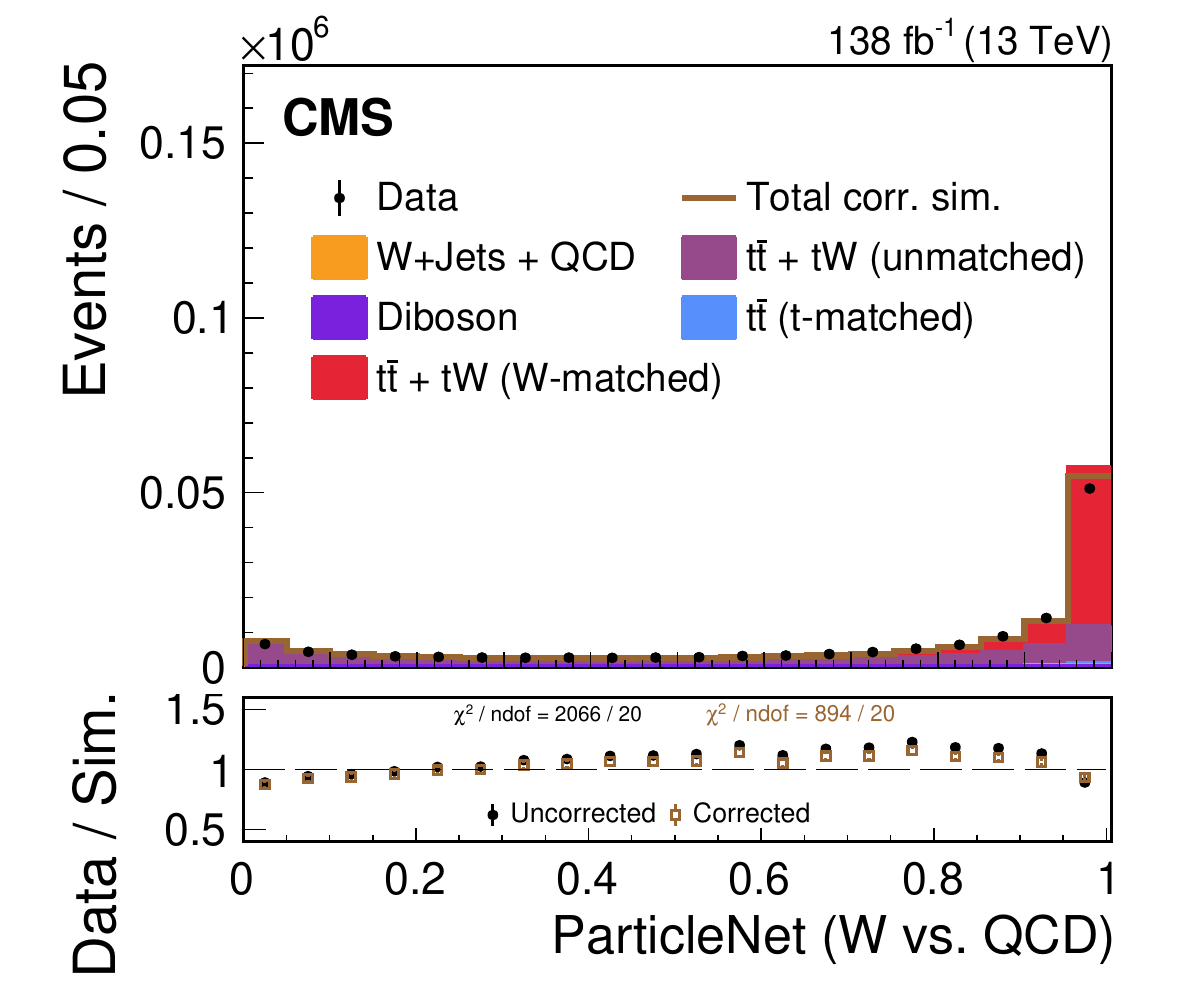} \\
  \includegraphics[width = 0.49\textwidth]{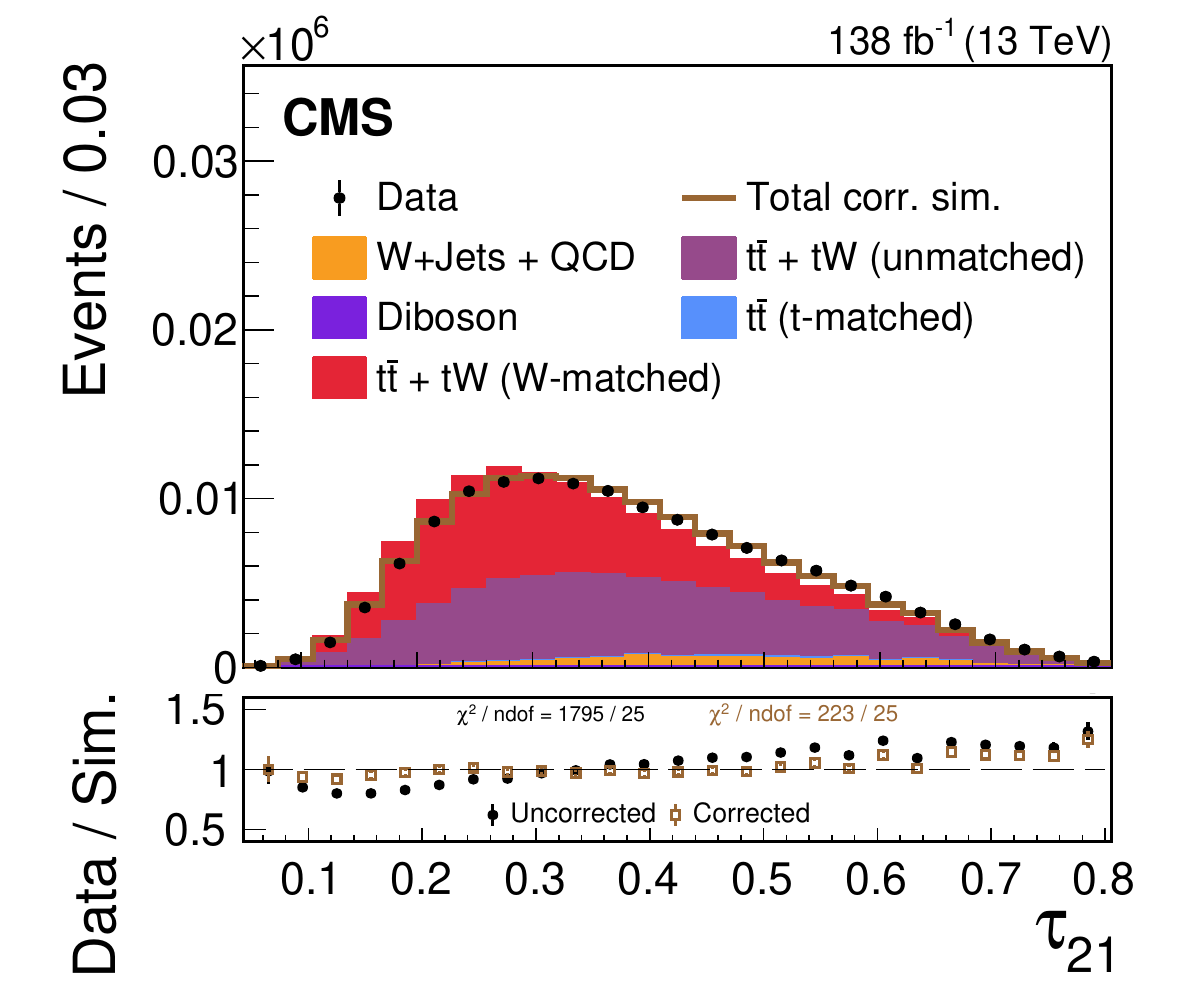}
  \includegraphics[width = 0.49\textwidth]{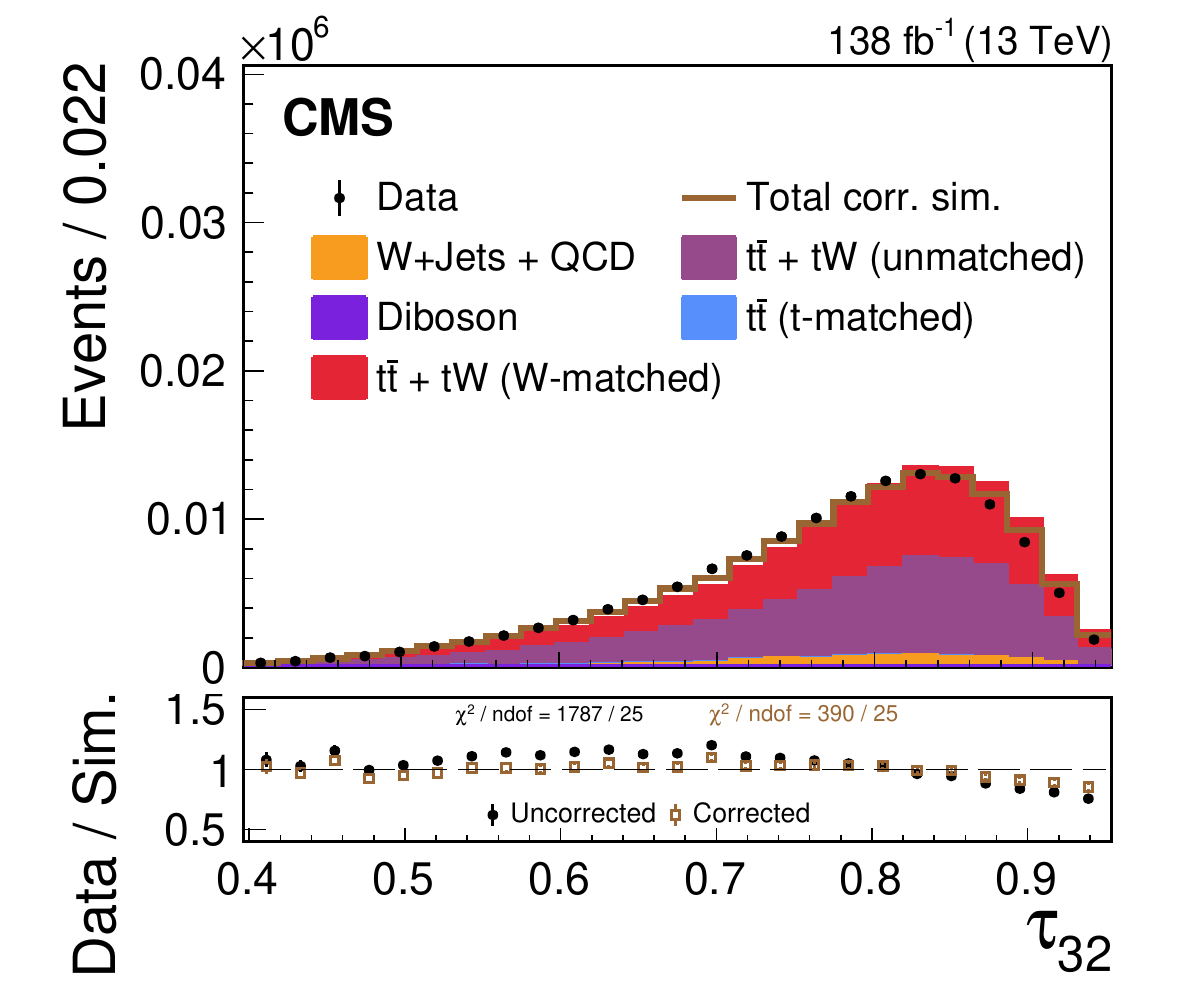} \\
  \caption{ 
    A comparison of the data-simulation agreement of various substructure observables in the \PW region.
    The distribution of various simulated processes, without the LJP correction applied, are shown in the colored histograms and observed data points are shown in black.
    The brown line shows the total simulated distribution after the LJP correction has been applied to the \PW-matched \ttbar and $\PQt\PW$ simulations;
    the other background processes are not corrected.
    Only statistical uncertainties are shown as vertical bars on the data points, and the computed $\chi^2$ is based only on statistical uncertainties.
    The black solid points (brown open boxes) in the lower panel show the ratio between the data and the total uncorrected (corrected) estimate from simulation.
    The data-simulation agreement of the various substructure distributions generally improves after applying the correction. 
}
  \label{fig:w_obs}
\end{figure}

\begin{figure}[!th]
  \centering
  \includegraphics[width = 0.49\textwidth]{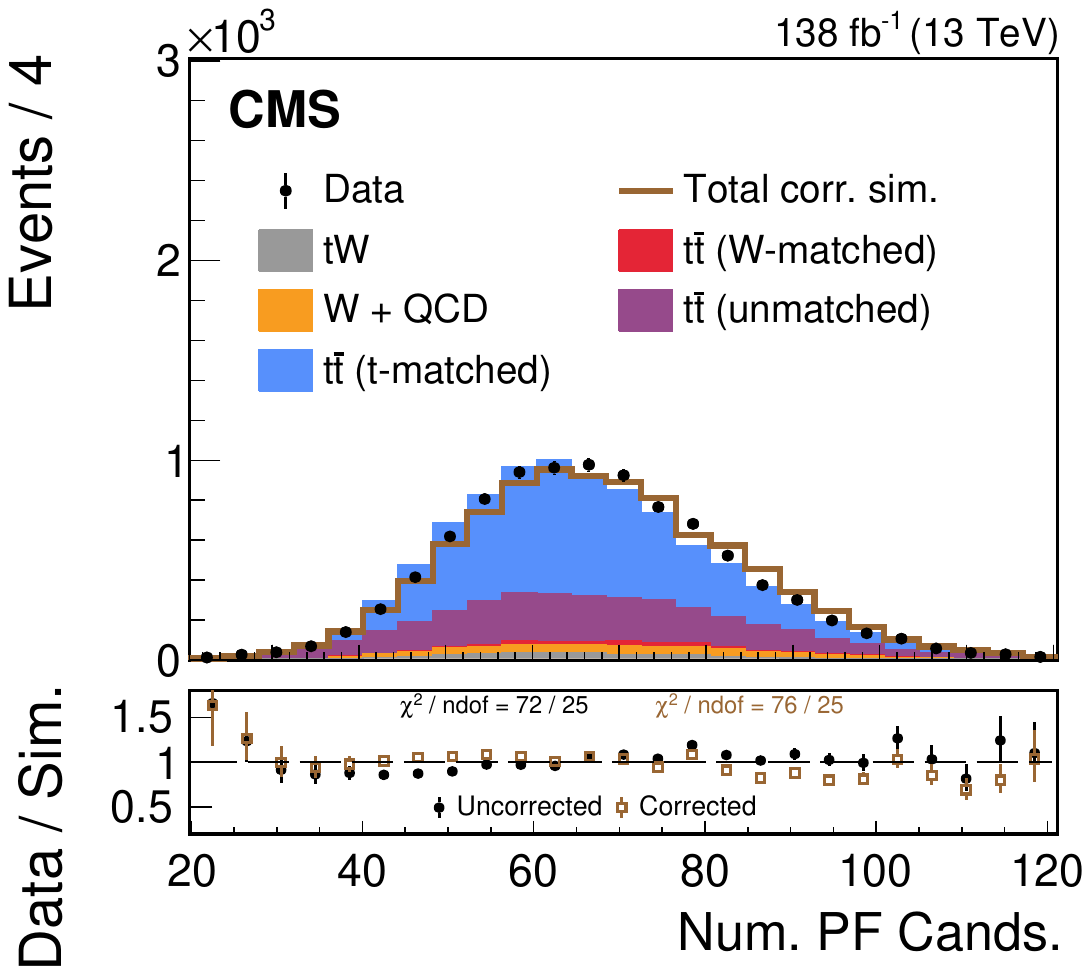} 
  \includegraphics[width = 0.49\textwidth]{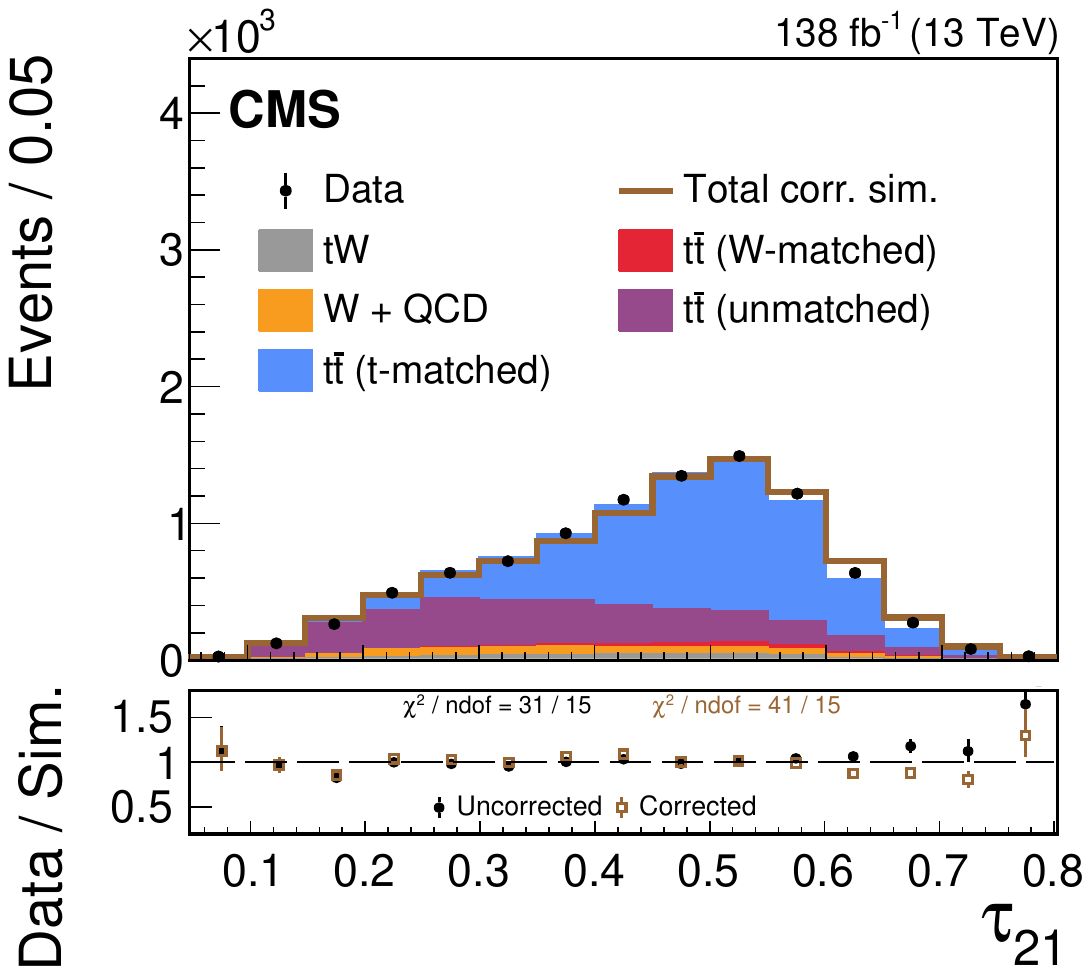} \\
  \includegraphics[width = 0.49\textwidth]{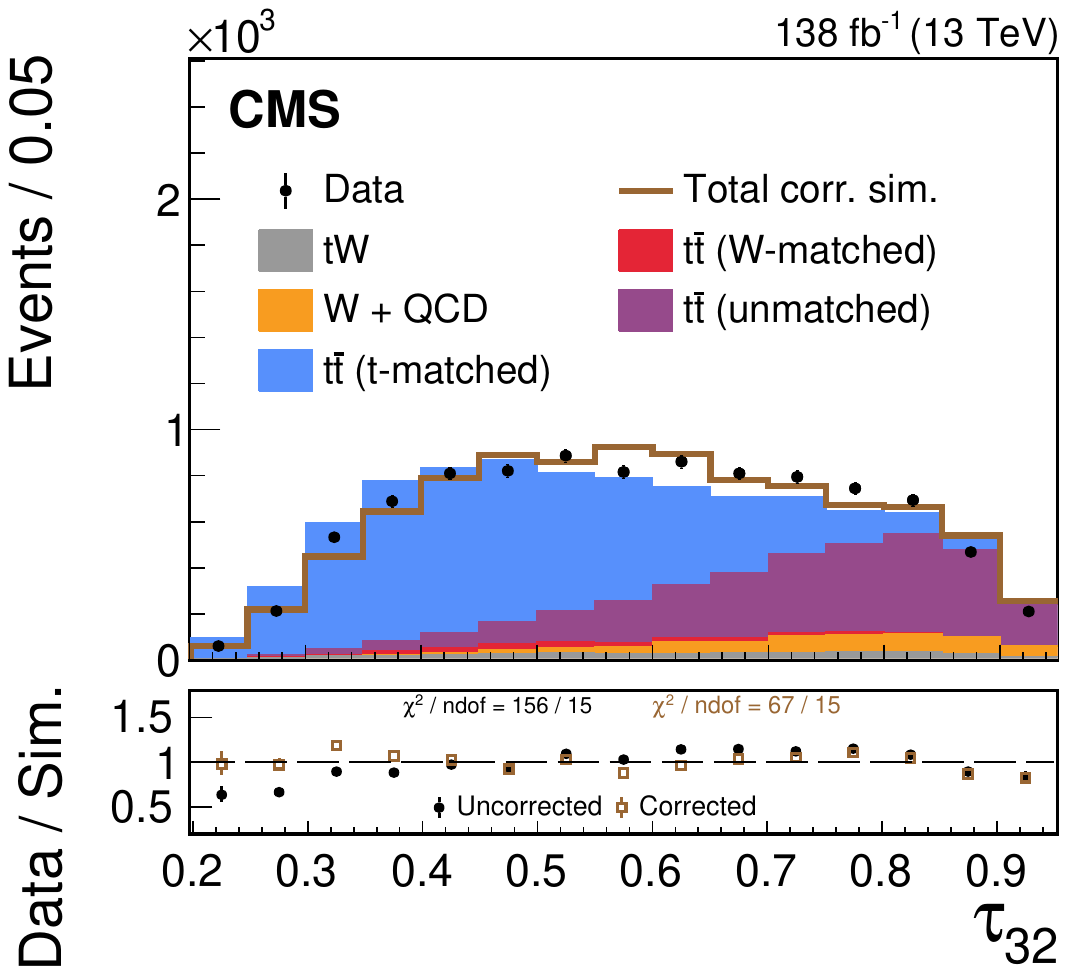} 
  \includegraphics[width = 0.49\textwidth]{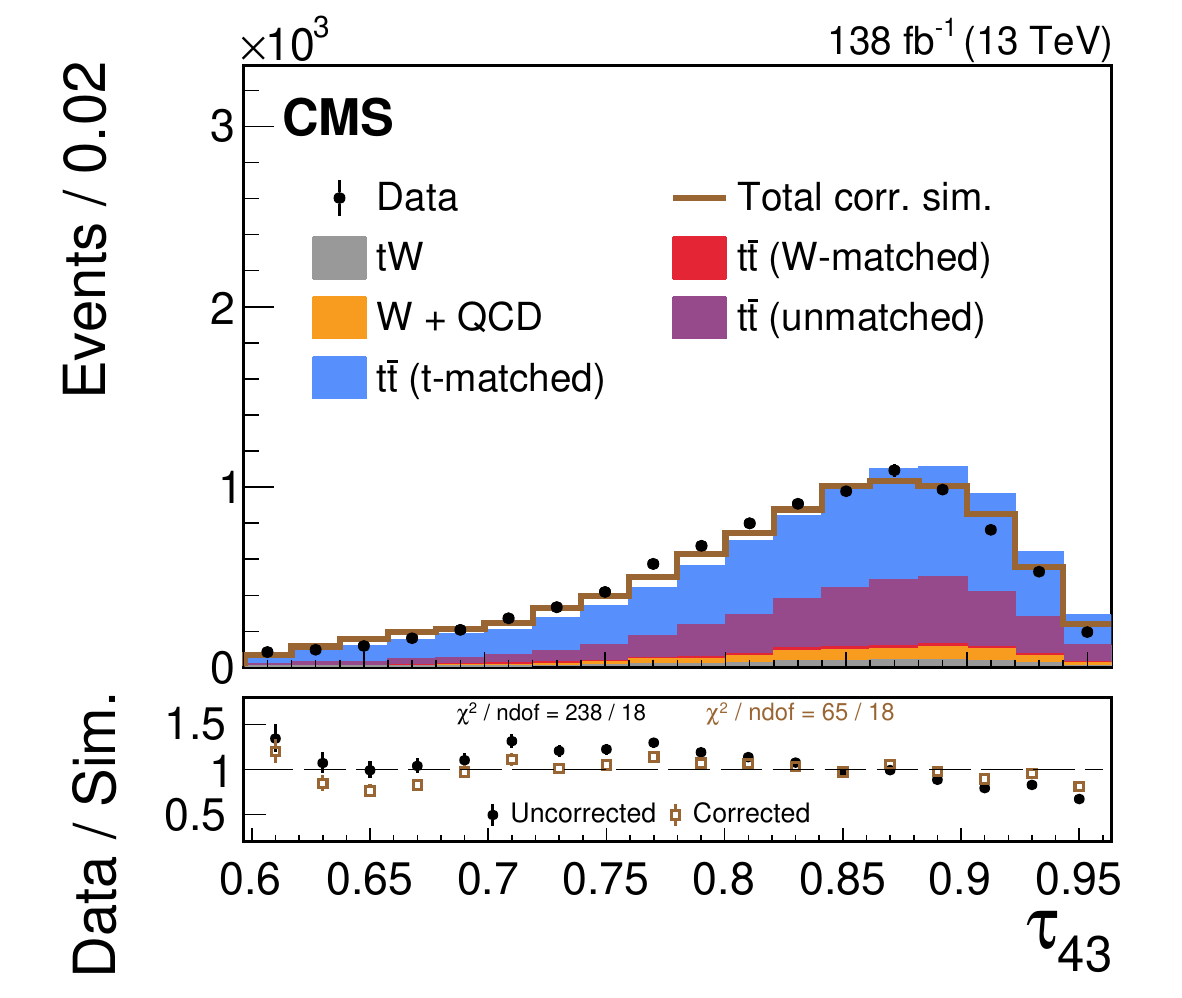}
  \caption{ 
    A comparison of the data-simulation agreement of various substructure observables in the \PQt region.
    The distribution of various simulated processes, without the LJP correction applied, are shown in the colored histograms and observed data points are shown in black.
    The brown line shows the total simulated distribution after the LJP correction has been applied to the \PQt-matched \ttbar simulation;
    the other background processes are not corrected.
    Only statistical uncertainties are shown as vertical bars on the data points, and the computed $\chi^2$ is based only on statistical uncertainties.
    The black solid points (brown open boxes) in the lower panel show the ratio between the data and the total uncorrected (corrected) estimate from simulation.
    The data-simulation agreement of the worst modeled substructure observables, $\tau_{32}$ and $\tau_{43}$, improves after applying the correction. 
}
  \label{fig:top_obs}
\end{figure}

As a first test, the correction procedure is applied to 2-pronged \PW jets.
The simulated samples of the $\PQt\PW$ process and \PW-matched \ttbar events are corrected following the above procedure.
The data-simulation agreement in the \PW region of various substructure observables before and after the correction are shown in Fig.~\ref{fig:w_obs}, including the \mSD, the $N$-subjetiness ratios~\cite{Thaler:2010tr} $\tau_{21}$ and $\tau_{32}$, and scores of the mass-decorrelated version of the \textsc{DeepAK8} tagger~\cite{CMS_tagging} and the \textsc{ParticleNet} tagger~\cite{Qu:2019gqs,CMS_ParticleNet}.
The correction improves the agreement of the substructure observables.
It should be noted that the correction is only applicable to a true multiprong jet arising from multiple quarks, and therefore is not applied to any of the background processes.
Therefore, it is expected that data and simulation still do not fully agree, though clear improvement is seen after the correction's application. 
In particular, the improved modeling of sophisticated deep neural network classifier scores, which are based on low-level features of the jet, indicates its wide applicability.
This demonstrates the ability of the method to correct multiple substructure observables simultaneously.

Next, the correction procedure is tested on 3-pronged jets in the \PQt region.
Distributions of various substructure observables before and after the correction are shown in Fig.~\ref{fig:top_obs}.
The correction significantly improvements the data-simulation agreement of the worst-modeled substructure observables ($\tau_{32}$ and $\tau_{43}$), demonstrating that the correction derived from 2-pronged \PW jets generalizes to a higher number of prongs.

\clearpage

\section{\texorpdfstring{\PYTHIA-\HERWIG}{PYTHIA-HERWIG} conversion study}
\label{sec:herwig_study}

The improved data-simulation agreement for \PW and top quark jets in Section~\ref{sec:apply_correction} is a powerful demonstration of the effectiveness of the method.
However, in that validation full closure between the data distribution and the corrected simulation is difficult to assess because of the presence of backgrounds which do not contain
true multiprong jets. 
Additionally, it does not validate the usage of the method on its main application target: jets with more than three prongs.

Therefore, to further validate the correction procedure and to test the extrapolation to jets with higher numbers of prongs, additional studies are done using simulation.
A simulation-based validation allows us to verify the performance in jets where the true distributions are known, 
and also allows us to study jets with higher prong multiplicities than present in data. 
This study also validates the uncertainties of the method, detailed in Section~\ref{sec:uncs}.

The studies use events showered with both \PYTHIA and \HERWIG, as discussed in Section~\ref{sec:samples}. 
A simulation-to-simulation correction is derived and is used to correct events showered with \PYTHIA to match the results from \HERWIG. 
The correction is applied to high-prong signals using \PYTHIA showering, and then compared with the same signals showered with \HERWIG; they found to agree within uncertainties.

The correction is derived in the same way as is done in data, as described in Section~\ref{sec:derive_correction}, except that \HERWIG-showered MC events are used in place of data.
The same selection criteria are used to select \PW jets from lepton+jets \ttbar events, and the same procedure is used to extract the \PYTHIA-to-\HERWIG LJP correction. 
The correction is tested on \PW jets and on BSM processes with four and six prongs.
The corrected \PYTHIA samples are then compared to the true \HERWIG sample to assess the performance.

Unmatched \ttbar decays, which constitute a background to the true \PW-matched \ttbar events, are included in the extraction of the LJP correction.
The \PYTHIA \ttbar sample is used to estimate and subtract this contribution before extracting the LJP density ratio. 
In the unmatched background the same uncertainties are used that were employed in the extraction from data, which is discussed in Section~\ref{sec:uncs}.
These uncertainties are propagated to the extracted LJP density ratio as systematic uncertainties.
Uncertainties due to limited MC event count are also included, but are small in most regions of the  LJP density. 
The LJP correction can also be extracted using only \PW-matched events from the \PYTHIA and \HERWIG samples, excluding the unmatched events entirely; negligible differences are observed in both the LJP density ratios and the corrected distributions, validating the treatment of the unmatched background. 

These extracted  LJP density ratios are then used to apply a correction to the \PYTHIA samples to better match the \HERWIG samples, including systematic uncertainties, as described in Section~\ref{sec:uncs}. 
The distributions of various substructure observables in the \PYTHIA samples are compared before and after the correction, using the true \HERWIG distributions as the reference. 
We assess the performance of the LJP correction on the calibration of deep neural network based jet taggers used widely in analyses, which may be sensitive to various features of the jet substructure. 
Specifically, we study the \textsc{DeepAK8} and \textsc{ParticleNet} tagger scores for \PW vs. QCD jet (\PW tag score) and $\PH\to\PW\PW\to 4\PQq$ vs. QCD jet ($\PH4\PQq$ score).

\begin{figure}[!t]h
  \centering
  \includegraphics[width = 0.49\textwidth]{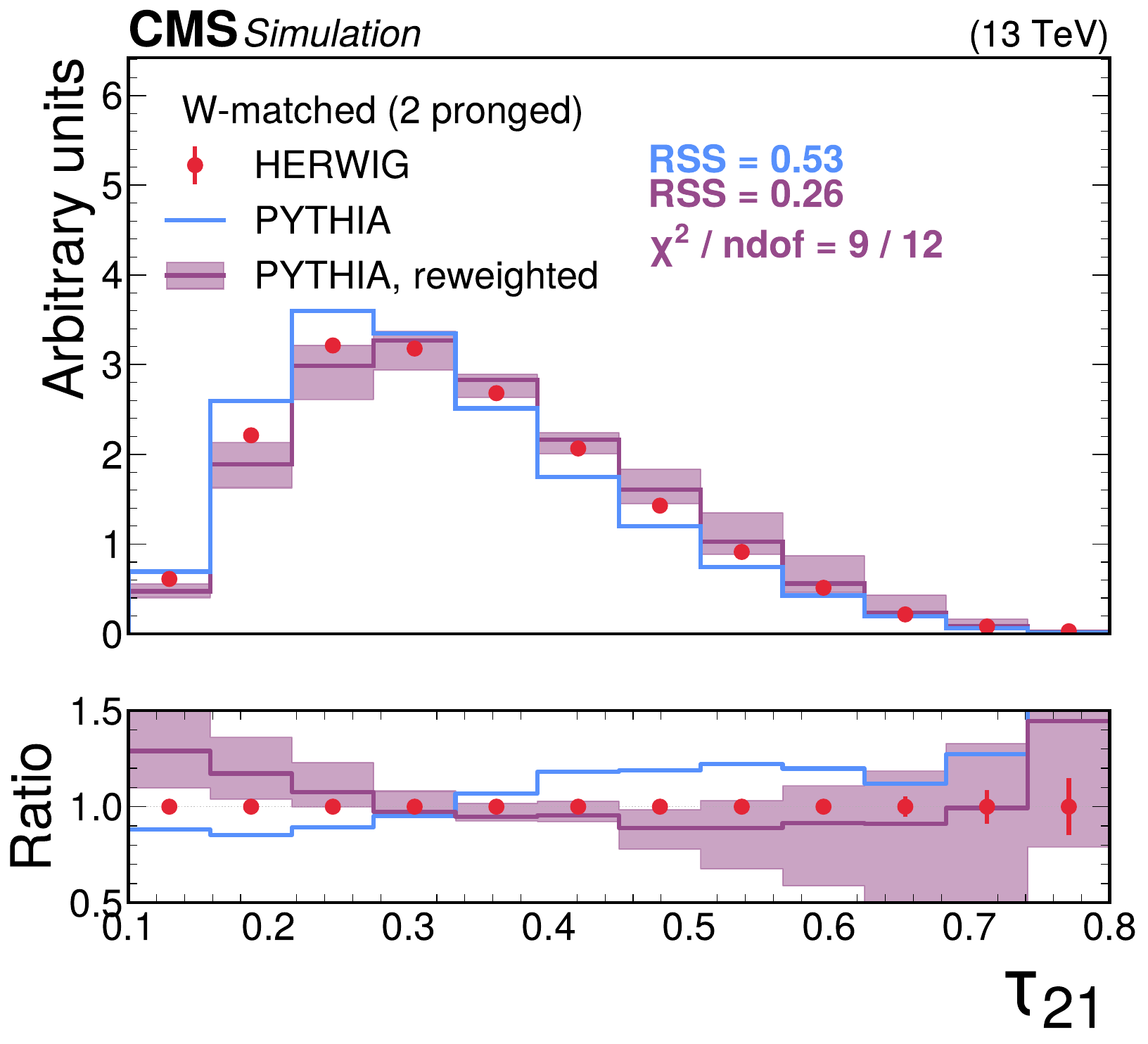} \\
  \includegraphics[width = 0.49\textwidth]{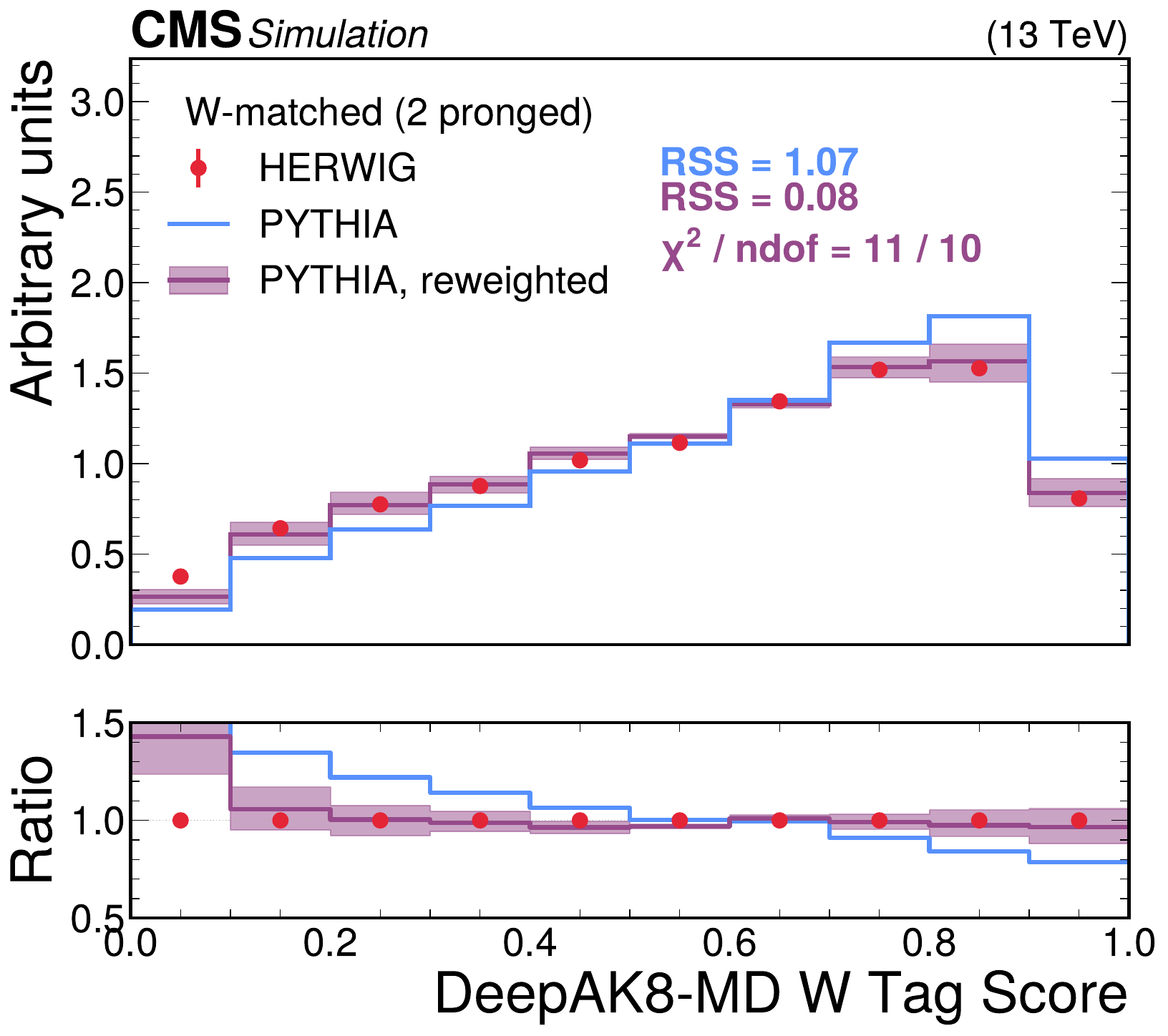}
  \includegraphics[width = 0.49\textwidth]{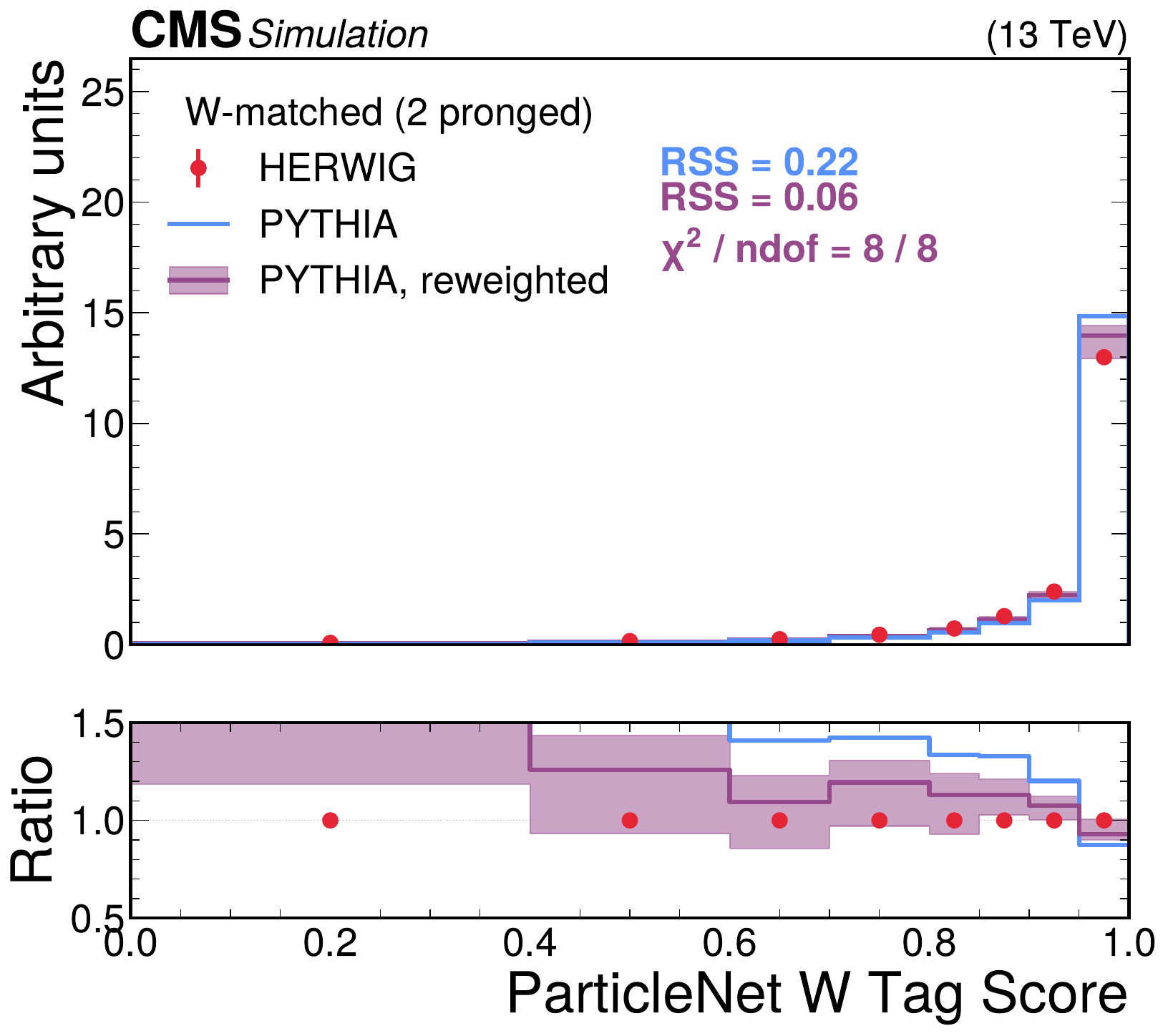}
  \caption{ 
    A comparison of the \HERWIG (red circles), \PYTHIA (blue lines) and reweighted \PYTHIA (purple lines) samples for \PW jets. 
    The systematic uncertainty in the reweighted \PYTHIA samples is shown in the light purple shading.
    The statistical uncertainty from the limited size of the simulated sample is shown as vertical red bars on the \HERWIG points.
    The lower panel shows the ratio of the two \PYTHIA distributions with respect to \HERWIG.
    The RSS between the \PYTHIA and \HERWIG samples is computed based on the squared difference in normalized bin yields.
    The $\chi^2$ value is computed using both the statistical uncertainties of the simulated samples and the systematic uncertainties
    in the correction procedure, and therefore assesses the full closure of the correction procedure.
    It is computed only for the reweighted \PYTHIA samples because the original sample does not have appropriate systematic uncertainties.
}
\label{fig:herwig_pythia_w}
\end{figure}

\begin{figure}[!th]
  \centering
  \includegraphics[width = 0.49\textwidth]{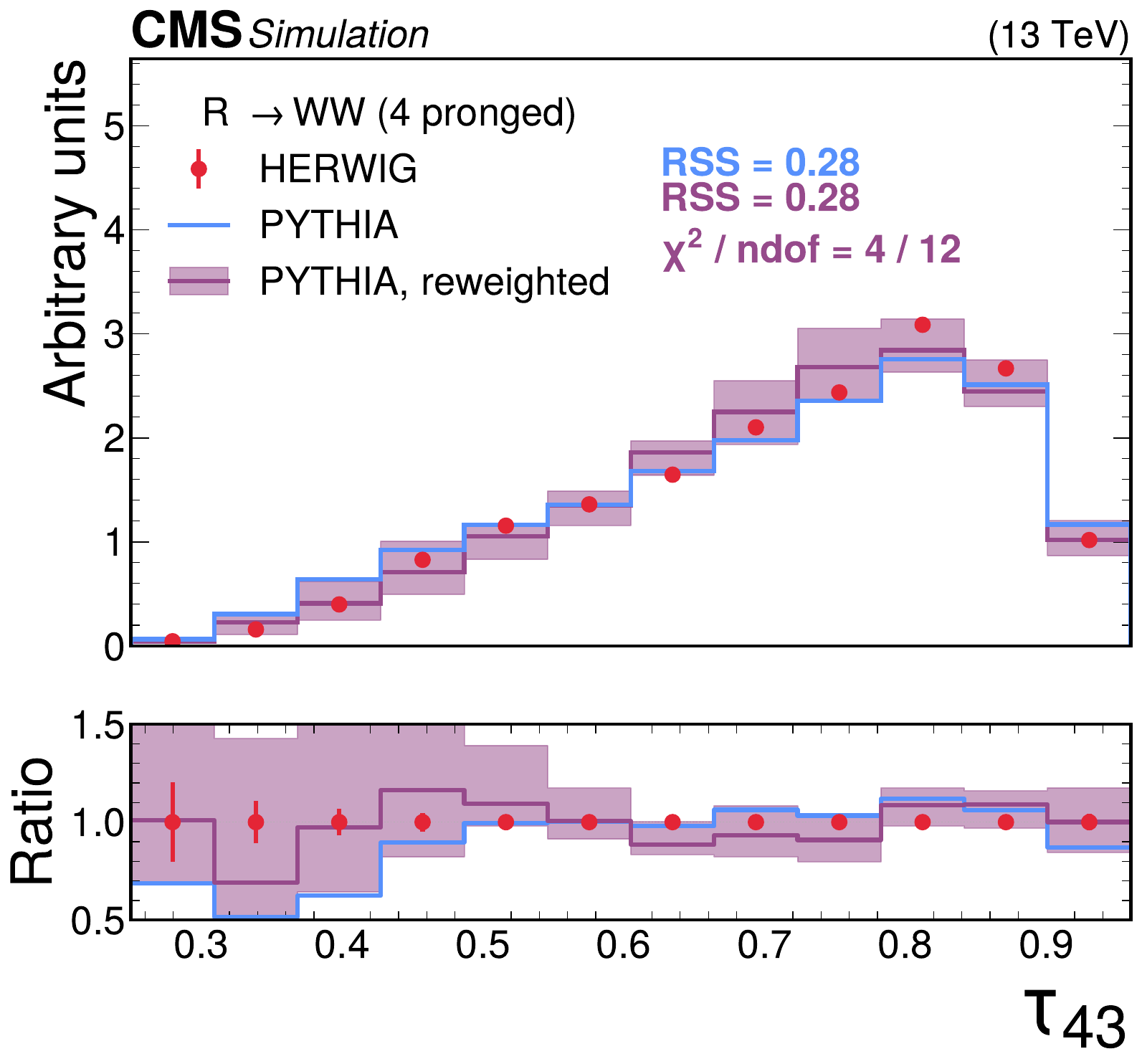} \\
  \includegraphics[width = 0.49\textwidth]{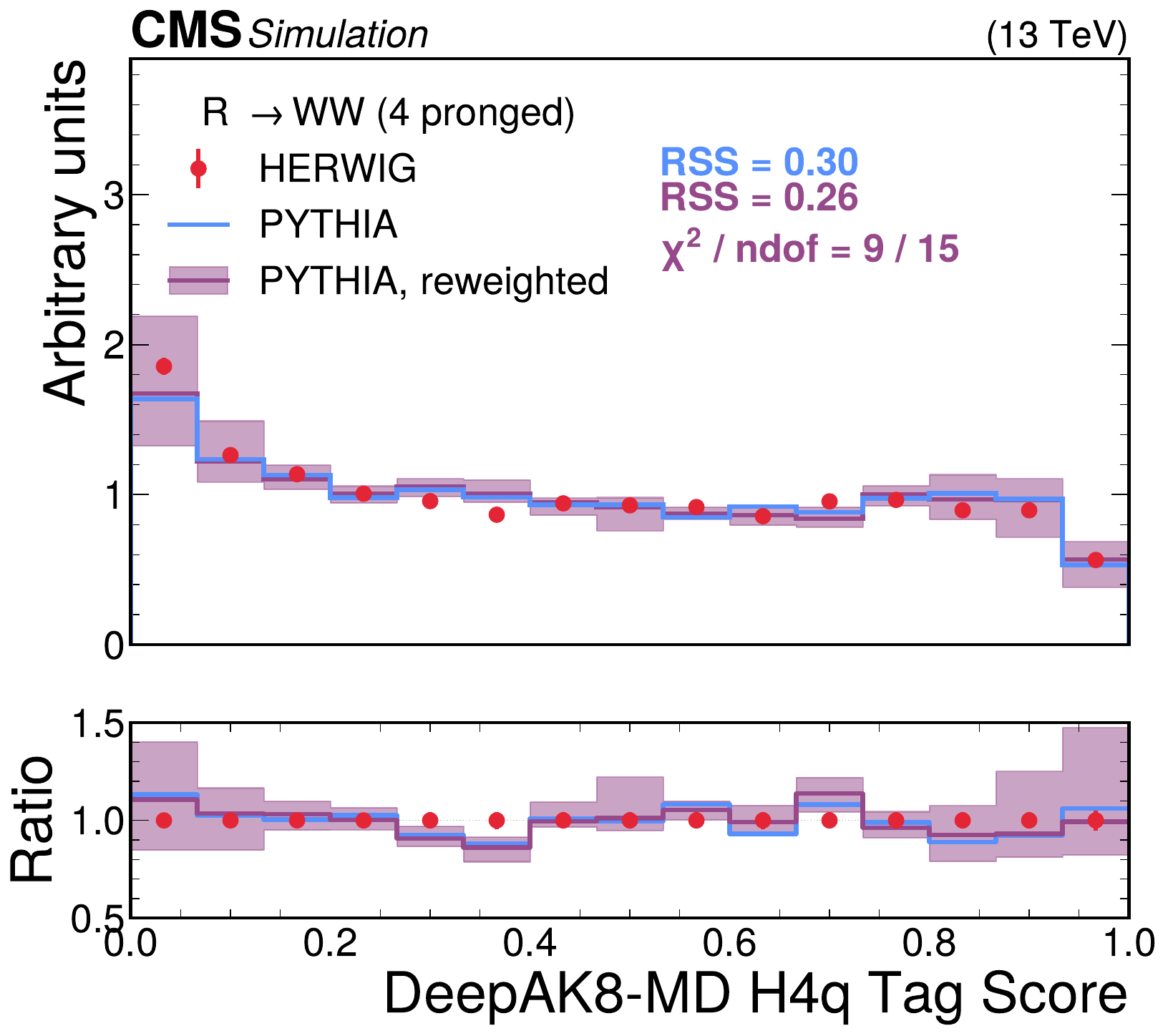}
  \includegraphics[width = 0.49\textwidth]{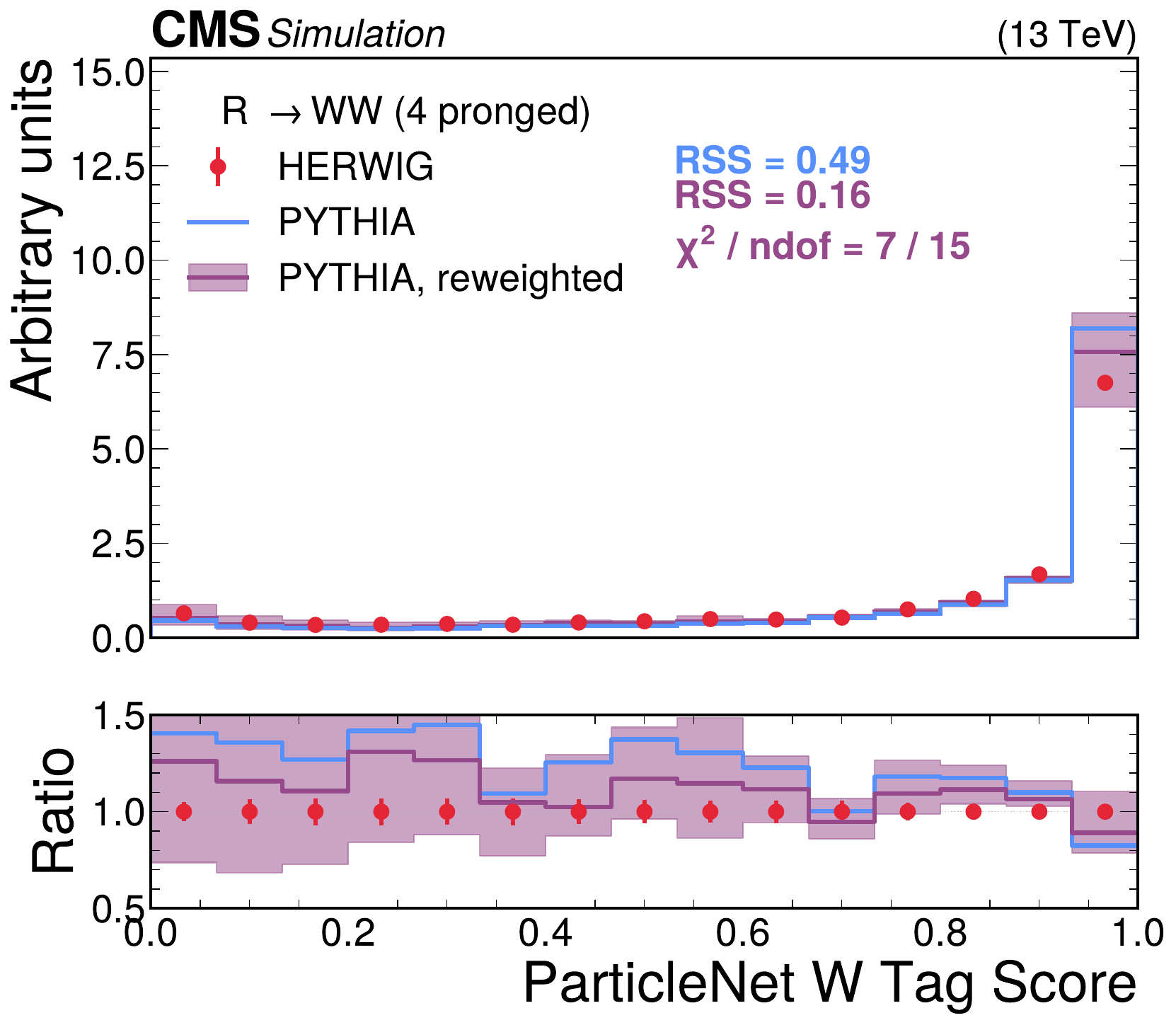}
  \caption{ 
    A comparison of the \HERWIG (red circles), \PYTHIA (blue lines) and reweighted \PYTHIA (purple lines) samples for $\PR \to \PW\PW \to 4\PQq$ jets. 
    The systematic uncertainty in the reweighted \PYTHIA samples is shown in the light purple shading.
    The statistical uncertainty from the limited size of the simulated sample is shown as vertical red bars on the \HERWIG points.
    The lower panel shows the ratio of the two \PYTHIA distributions with respect to \HERWIG.
    The RSS between the \PYTHIA and \HERWIG samples is computed based on the squared difference in normalized bin yields.
    The $\chi^2$ value is computed using both the statistical uncertainties of the simulated samples and the systematic uncertainties
    in the correction procedure, and therefore assesses the full closure of the correction procedure.
    It is computed only for the reweighted \PYTHIA samples because the original sample does not have appropriate systematic uncertainties.
}
\label{fig:herwig_pythia_Wkk}
\end{figure}

\begin{figure}[!th]
  \centering
  \includegraphics[width = 0.49\textwidth]{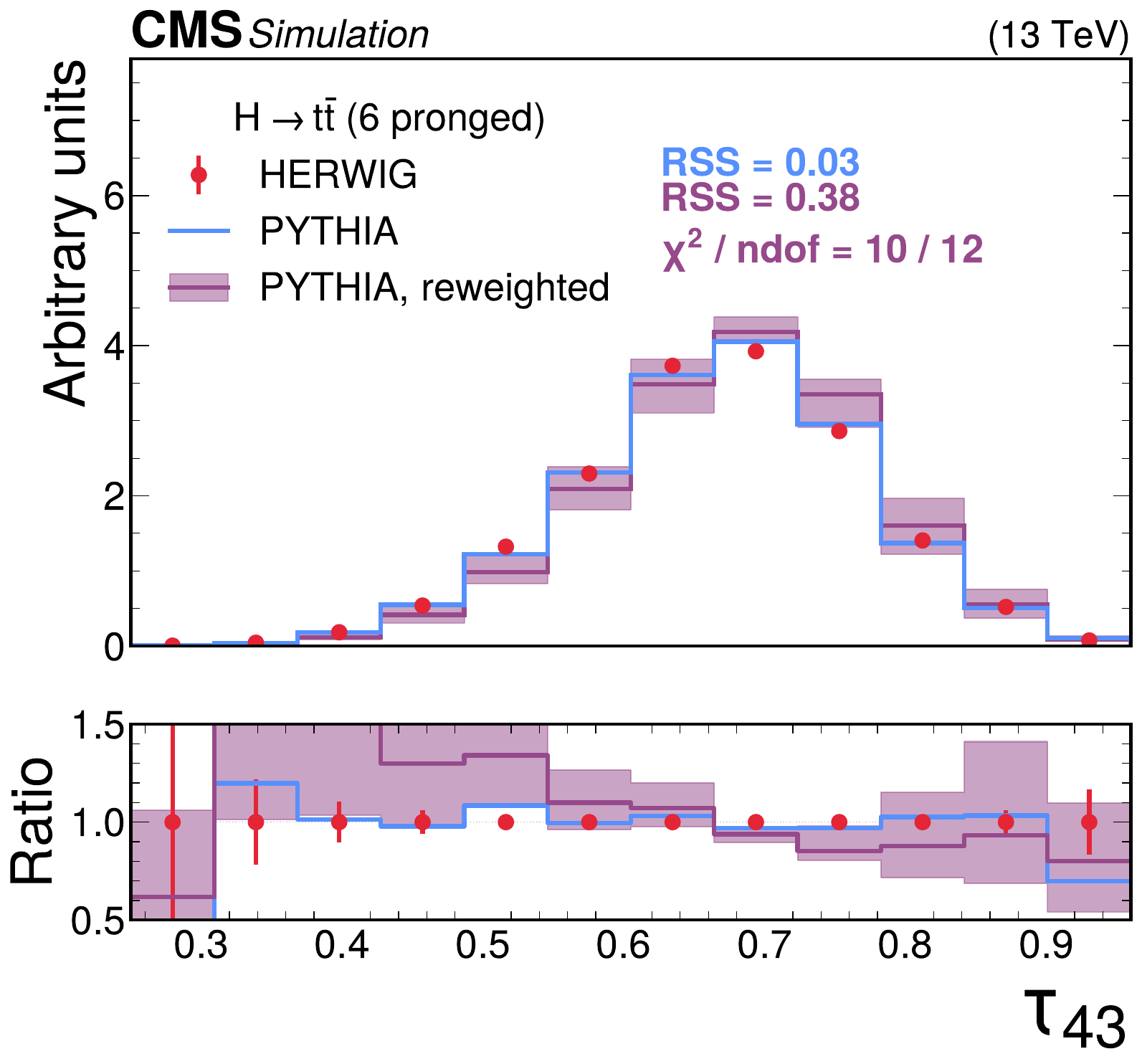} \\
  \includegraphics[width = 0.49\textwidth]{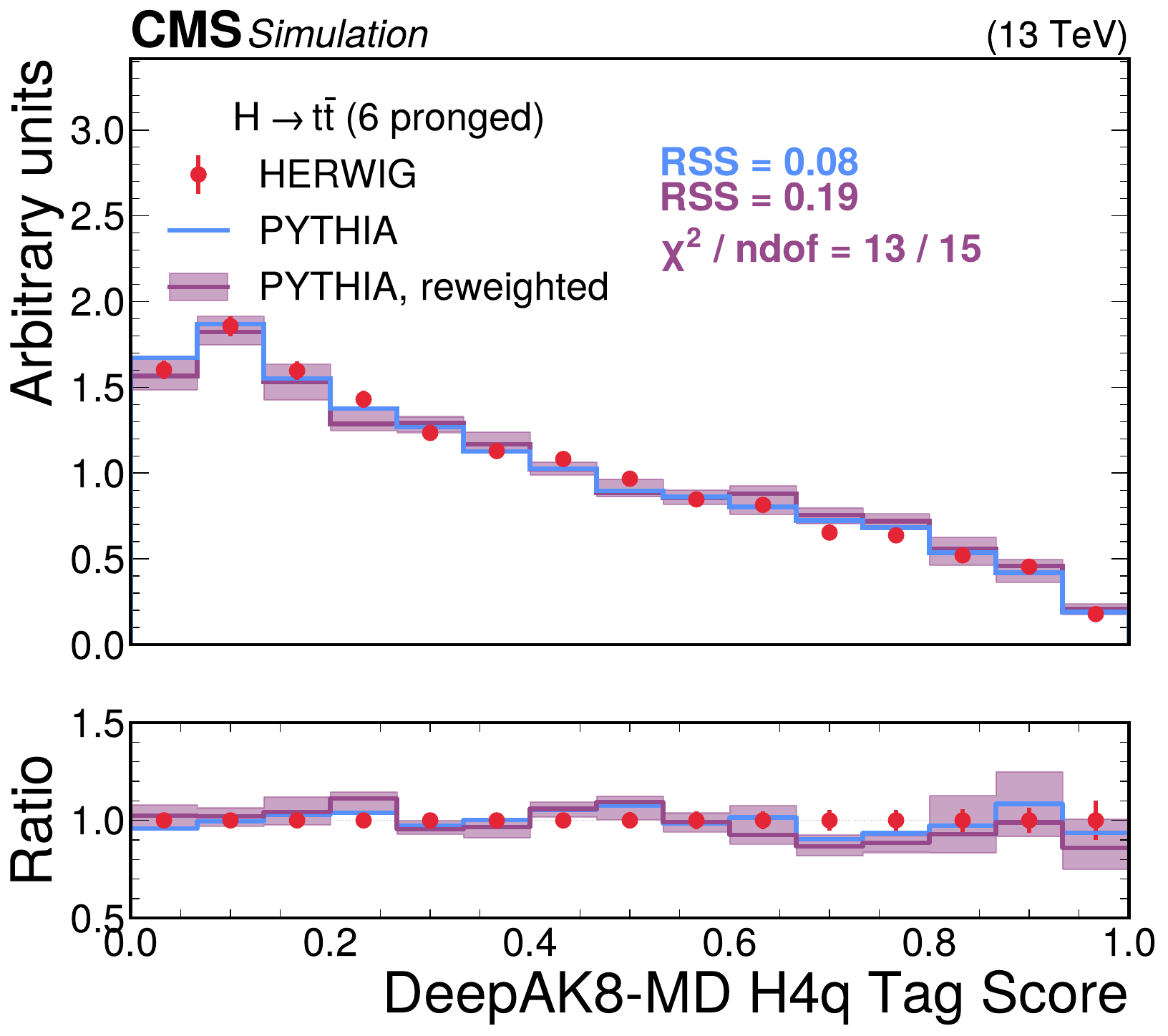}
  \includegraphics[width = 0.49\textwidth]{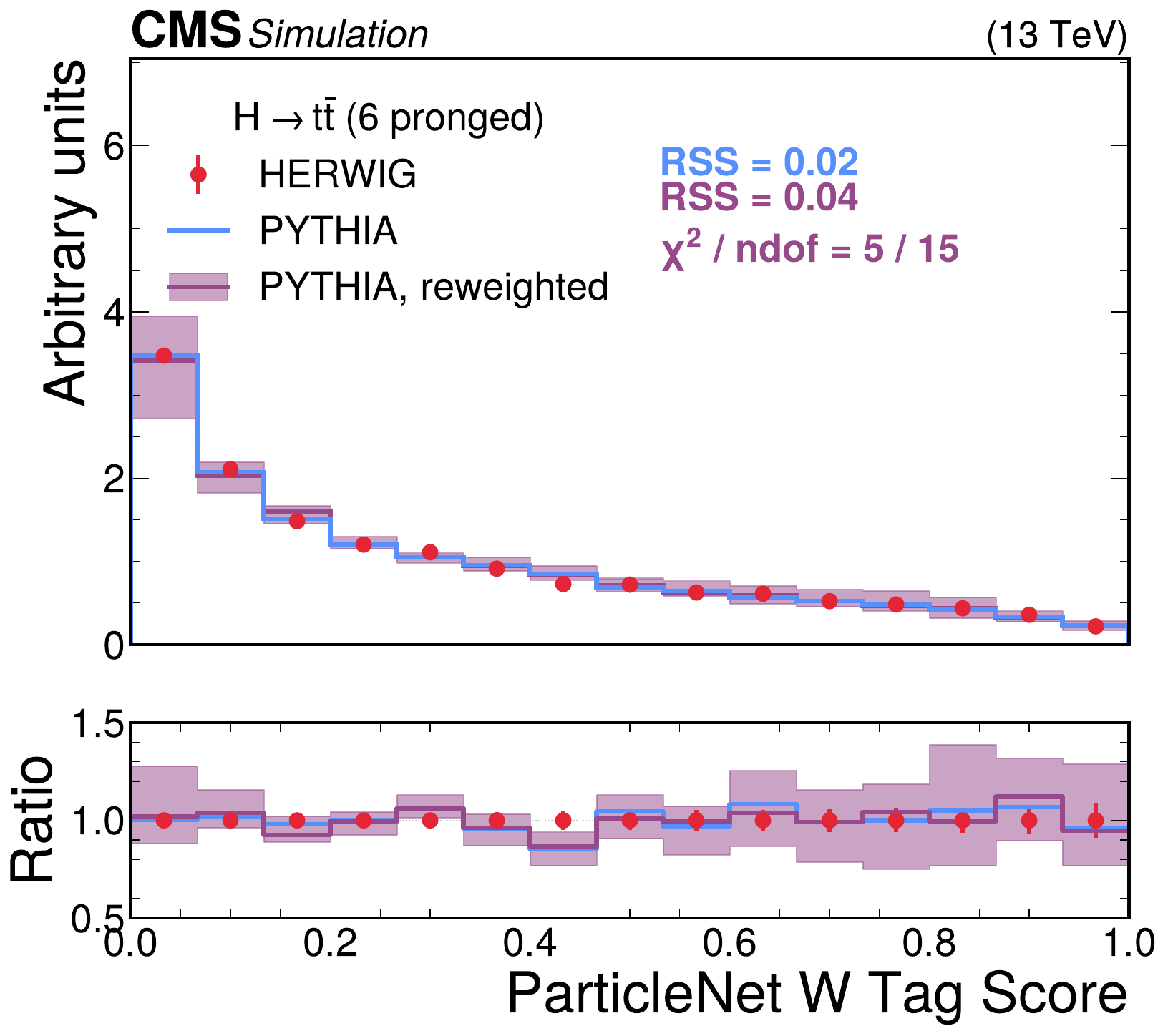}
  \caption{ 
  A comparison of the \HERWIG (red circles), \PYTHIA (blue lines) and reweighted \PYTHIA (purple lines) samples for $\PH \to \ttbar \to 6\PQq$ jets.
    The systematic uncertainty in the reweighted \PYTHIA samples is shown in the light purple shading.
    The statistical uncertainty from the limited size of the simulated sample is shown as vertical red bars on the \HERWIG points.
    The lower panel shows the ratio of the two \PYTHIA distributions with respect to \HERWIG.
    The RSS between the \PYTHIA and \HERWIG samples is computed based on the squared difference in normalized bin yields.
    The $\chi^2$ value is computed using both the statistical uncertainties of the simulated samples and the systematic uncertainties
    in the correction procedure, and therefore assesses the full closure of the correction procedure.
    It is computed only for the reweighted \PYTHIA samples because the original sample does not have appropriate systematic uncertainties.
}
\label{fig:herwig_pythia_YtoHH}
\end{figure}

The performance of the \PYTHIA-to-\HERWIG correction is checked across several different types of jets. 
The agreement between the \PYTHIA and \HERWIG distributions before and after the correction is quantified with the residual sum of squares (RSS), which sums the squared difference between the bin yields of the two distributions. 
A $\chi^2$ is also computed to assess closure between the corrected \PYTHIA and true \HERWIG samples, which takes into account the systematic uncertainties in the \PYTHIA distribution. 
Results for 2-pronged \PW jets are shown in Fig.~\ref{fig:herwig_pythia_w}.
Results for 4-pronged jets from the $\PR \to \PW\PW \to 4\PQq$ process are shown in Fig.~\ref{fig:herwig_pythia_Wkk}.
Results on 6-pronged jets due to boosted $\PH \to \ttbar \to 6\PQq$ decays are shown in Fig.~\ref{fig:herwig_pythia_YtoHH}.
A comparison of tagging efficiencies for several jet types is given in Table \ref{tab:herwig_cut_effs}.

For \PW jets, the correction works well for observables related to \PW tagging. 
The agreement between \PYTHIA and \HERWIG for typical substructure observables used to tag \PW jets, including $\tau_{21}$, the \textsc{ParticleNet} \PW vs. QCD score, and \textsc{DeepAK8-MD} \PW vs. QCD score, improves significantly with the correction.
The systematic uncertainties cover residual differences between the reweighted \PYTHIA and the true \HERWIG simulation.

For the $\PR \to \PW\PW \to 4\PQq$ and $\PH \to \ttbar$ signals, there is already good agreement between \PYTHIA and \HERWIG, so 
we see limited improvements in RSS values.
For the six-pronged jets from $\PH \to \ttbar$, there is a slight overcorrection, leading to slightly worse RSS values.
The systematic uncertainties again cover residual differences in all cases.

\begin{table}[!th]
    \centering
    \topcaption{A comparison of the tagging efficiency in the nominal \PYTHIA simulation, the corrected \PYTHIA simulation and the \HERWIG simulation for jets of various kinds. 
      Uncertainties in the correction procedure are propagated to evaluate the uncertainty in the tagging efficiency in the corrected \PYTHIA simulation.
     Details are given in the text.}
      \cmsTable{
    \begin{tabular}{ c c c c c c c }
      Jet type (prongs) & Selection & Nom. \PYTHIA eff. & Corrected \PYTHIA eff. & \HERWIG eff.  \\
      \hline
      \PW (2) & $\tau_{21} < 0.3$ & 0.43 & $0.33^{+0.04}_{-0.06}$ & 0.36 \\
      top (3) &  $\tau_{32} < 0.5$ & 0.49 & $0.39^{+0.08}_{-0.10}$ & 0.42 \\
      $\PR \to \PW\PW$ (4) & DeepAK8-MD H4q $> 0.4$  & 0.53 & $0.52 \pm 0.03 $ & 0.53\\
      $\PH \to \ttbar$ (6) & DeepAK8-MD H4q $> 0.4$  & 0.41 & $0.42 \pm 0.02 $ & 0.41\\
    \end{tabular}
  }
  \label{tab:herwig_cut_effs}
\end{table}

These results demonstrate that the correction derived from a two-pronged jet can successfully extrapolate to higher pronged jets.
The LJP correction is also shown to improve the modeling of sophisticated ML-based tagging scores. 
Further, the systematic uncertainties cover residual differences between the true and corrected distributions, even for up to 6-pronged jets. 
This justifies the usage of the LJP method for calibrating substructure observables for jets with a high number of prongs.

\section{Uncertainties}
\label{sec:uncs}

There are various sources of uncertainty in the correction procedure, including statistical uncertainties in the derived LJP density ratio, systematic uncertainties in the derived LJP density ratio, and systematic uncertainties relating to the application of the procedure to a given signal process. 

\subsection{Statistical uncertainty in data/simulation ratio}
The core aspect of the correction procedure is the data/simulation subjet LJP density ratio that is derived from the sample of \PW jets. 
The finite number of data events in this sample leads to a statistical uncertainty in the data/simulation ratio in each bin of the LJP density.
The uncertainty from limited MC statistics is also considered and propagated to the uncertainty in the ratio, but is negligible. 

The uncorrelated statistical uncertainty in each LJP density ratio bin should be propagated into the uncertainty in the event-level correction. 
To assign an appropriate statistical uncertainty, 100 variations of the LJP density ratio are constructed. 
Each variation is based on an independent Gaussian smearing of the central value in each bin according to the bin's statistical uncertainty, constrained so that the ratio is always positive.
The correction is recomputed for these 100 variations. 
For any computed quantity, the standard deviation of the resultant quantities based the smeared variations is taken as the statistical uncertainty in the nominal correction.
It was checked that the results of the procedure are insensitive to the number of variations.

\subsection{Systematic uncertainty in data/simulation ratio}
There are additional systematic uncertainties on the ratio stemming from uncertainties in the simulation used for estimation of the background processes that are subtracted from the data, and in the simulation that constitutes the denominator of the ratio.
The uncertainties in the background estimation are generally dominant.
For a given source of systematic uncertainty, the relevant simulated samples are reweighted according the up and down variations of the uncertainty, and the corresponding systematic variation of the ratio is derived.
Some sources of uncertainty are applied simultaneously to the signal (\PW-matched \ttbar and $\PQt\PW$) and background simulations, whereas some are applied only to the backgrounds.
This is because the parameters of the simulation used in the denominator of the ratio must be consistent with the parameters of the simulation to be reweighted, and, therefore, we do not apply the variations for uncertainties that would change the substructure observables of the signal process simulation. 

Several sources of uncertainty in the background simulation are included. 
The largest source of uncertainty comes from the modeling of the parton shower, which is estimated by varying the modeling of initial- and final-state radiation in the \PYTHIA simulation.
Uncertainties in the modeling of pileup are evaluated by varying the inelastic $\Pp\Pp$ cross section used for the pileup simulation. 
Each background process is also assigned a normalization uncertainty based on the uncertainty in their cross section. 
The $\PQt\PW$ and the $s$- and $t$-channel single top quark processes are assigned a 5\% uncertainty~\cite{Campbell:2020fhf,PDF4LHCWorkingGroup:2022cjn}. 
Diboson processes are assigned a 4\% uncertainty~\cite{Gehrmann:2014fva,Cascioli:2014yka,Campbell:2011bn}.
The $\PW+\text{jets}$ and QCD samples are assigned a 10\% uncertainty. 
The unmatched \ttbar component is given a 6\% uncertainty, based on the difference in the unmatched fraction between \PYTHIA and \HERWIG samples.
The normalization uncertainty in each background is assumed to be independent.

The set of uncertainties applied simultaneously to the signal and background MC samples are as follows.
The uncertainty in the PDF modeling is obtained for each event based on the 100 NNPDF replicas~\cite{NNPDF:2017mvq}.
Three sets of variations of the renormalization and factorization scales are considered: one where only the renormalization scale is varied, one where the factorization scale is varied, and  one where both are simultaneously varied in the same direction. We do not consider the unphysical variation where the two scales are varied in opposite directions.
The uncertainty in the \PQb tagging efficiency is assessed by varying data vs. simulation \PQb tagging efficiency correction factors within their uncertainties.
Uncertainties in the muon identification, isolation,  and trigger efficiency are also considered.
This is done separately for the identification uncertainty ($\approx$0.1\%), isolation uncertainty ($\approx$0.1\%), and trigger uncertainty ($\approx$0.2\%).
Uncertainties in energy scale and resolution of simulated jets can affect the subjet \pt bin of the LJP density that splittings are placed into.
Jet momenta are varied up and down according to their uncertainties and used to construct systematic variations of the LJP density.

For computational efficiency, the contribution from these sources of uncertainty in the LJP density ratio are combined into one. 
For each bin of the LJP density ratio, the variation in the ratio for each uncertainty is computed, and then the variations are added in quadrature to obtain a single overall systematic uncertainty. 
The largest source of systematic uncertainty is the modeling of the parton shower for the background processes; nevertheless, statistical uncertainties are dominant in nearly all bins of the LJP density ratio.
The \pt extrapolation procedure is also applied to systematic variations of the LJP density ratio.
In the application of the method, the correction procedure is repeated using these overall systematic variations of the LJP density ratio. 
The resulting variations in the weights encode the uncertainty in the correction due to the systematic uncertainty in the LJP density ratio.
It was checked that the uncertainty in the correction for \PW tagging derived using a single overall variation for the systematic uncertainties is similar to that derived from treating each source of uncertainty individually and combining the results in quadrature.

\subsection{Uncertainty from the \texorpdfstring{\pt}{pt} extrapolation} 
\label{subsec:pt_unc}

As discussed in Section \ref{subsec:pt_extrap}, the LJP correction is extrapolated to accommodate subjets with \pt greater than that of subjets in the \PW jet sample. 
The uncertainty in the \pt extrapolation is determined from the uncertainties in the fit parameters.
Because the fit in each LJP density bin is done independently, the uncertainties in each bin are uncorrelated. 
Additionally, since independently varying the fit parameters one by one in each bin is computationally infeasible, we adopt a similar procedure as for the statistical uncertainties.
First, 100 variations of the extrapolated LJP density ratios are constructed by randomly varying the fit parameters according to their fitted uncertainties, accounting for any correlations. 
In the application of the LJP correction, the corrected quantity is computed with these 100 variations, and the standard deviation is taken as the uncertainty due to the \pt extrapolation. 

\subsection{Uncertainty from \texorpdfstring{\PQb}{b} quark initiated subjets} 

It is known that heavier quarks shower differently from lighter quarks~\cite{deadcone_theory, ALICE_deadcone}. 
Therefore, the correction derived from the sample of \PW jets, which is flavor-agnostic and consists primarily of decays to lighter quarks,
may not serve as a suitable correction for subjets initiated by heavier quarks.
Subjets coming from \PQb quarks in particular would be expected to be the most different, and form a key component of some multiprong jets, \eg, top quark jets.
This difference may be mitigated somewhat by the fact that our application focuses on large-$R$ jets, which generally contain subjets with \pt significantly larger than the \PQb quark mass, so mass-related effects will be reduced for the initial stages of the shower~\cite{deadcone_theory}.

To assess the uncertainty in the application of the LJP correction to \PQb quark subjets, we compare the LJP density distributions of \PQb quarks and light quarks in simulation.
From a simulated sample of boosted top quarks, we recluster the top quark jets into three subjets, and then separately construct the LJP densities of subjets from \PQb quarks and from light quarks originating in the \PW boson decays.
Of course a small fraction of $\PW\to\PQc\PQb$ decays are included in the light quark sample, since they are present in the observed events.
The average difference between the LJP densities is found to be $\sim$5\%. 

The ratio of these two densities is a measure of the difference in the showering of these two different kinds of subjets, and is referred to as the \PQb/light LJP density ratio.
The \PQb/light ratio is used as an estimate of the uncertainty in the \PW-derived correction when applied to a signal with \PQb quark subjets. 
Specifically, for simulated signal processes which produce \PQb quark subjets, generator-level information is used to identify which of 
the reclustered subjets are matched to \PQb quarks by using a $\Delta R < 0.2$ criterion. 
The systematic uncertainty in the corrected observable is evaluated by performing the correction with the nominal LJP density data-simulation ratio scaled by the \PQb/light LJP density ratio.

\subsection{Number of prongs uncertainty}
The choice of how many subjets the jet should be reclustered into is done on a case-by-case basis to match the number of prongs inside the AK8 jet.
Specifically, the number of subjets used for the reclustering is set to the number of generator-level quarks from the hard process that are within $\Delta R < 0.8$ of the jet. 
However, there can be ambiguity in the number of prongs. 
In such cases, an uncertainty in the number of prongs is evaluated by varying the number of subjets specified in the reclustering with the exclusive \kt algorithm, and evaluating the LJP correction with an alternative set of subjets.

Two cases are considered. 
First, a quark near the AK8 jet boundary will possibly have some radiation both inside and outside the cone of the jet. 
For any jet in which a quark is located within $0.7 < \Delta R < 0.8$ ($0.8 < \Delta R < 0.9$) of the jet, the reclustering is performed with one fewer (one additional) subjet. 
Second, if a single subjet is the closest subjet to two generator-level quarks, both within $\Delta R<0.2$, or if there is a generator-level quark not matched to a subjet, the reclustering is performed with both one additional and one fewer subjet. 
Regardless of the number of subjets used in the exclusive \kt algorithm, only subjets within $\Delta R<0.2$ of a generator-level quark are used in the correction.
Quarks that are still not matched after varying the number of subjets are referred to as `unclustered', and are assigned a dedicated systematic uncertainty described below. 

\subsection{Unclustered prongs uncertainty}
If a generator-level quark is not matched to a subjet, in both the nominal exclusive \kt reclustering and in the variations in the number of prongs, it is considered `unclustered'. 
These unclustered quarks are not corrected by the procedure, and, therefore, a conservative approach is taken to assess their impact on jet substructure observables. 
The systematic uncertainty due to unclustered quarks is defined by varying the weights of events containing unclustered quarks up and down by a factor of five, the maximum reweighting factor considered in the correction. 
Events that do not contain any unclustered quarks retain their nominal weights. 

We note that because the LJP correction weights are normalized such that their average value is one, this uncertainty is meaningful only if unclustered quarks are present in a subdominant fraction ($\lesssim$50\%) of the total events. 
In the limit that every event in the sample has an unclustered quark, all event weights would be varied by the same factor, and the normalization procedure would effectively undo the variation. 
Therefore, we check that, in the various simulated samples, the fraction of events containing unclustered quarks is subdominant.

\begin{figure}[!th]
  \centering
  \includegraphics[width = 0.49\textwidth]{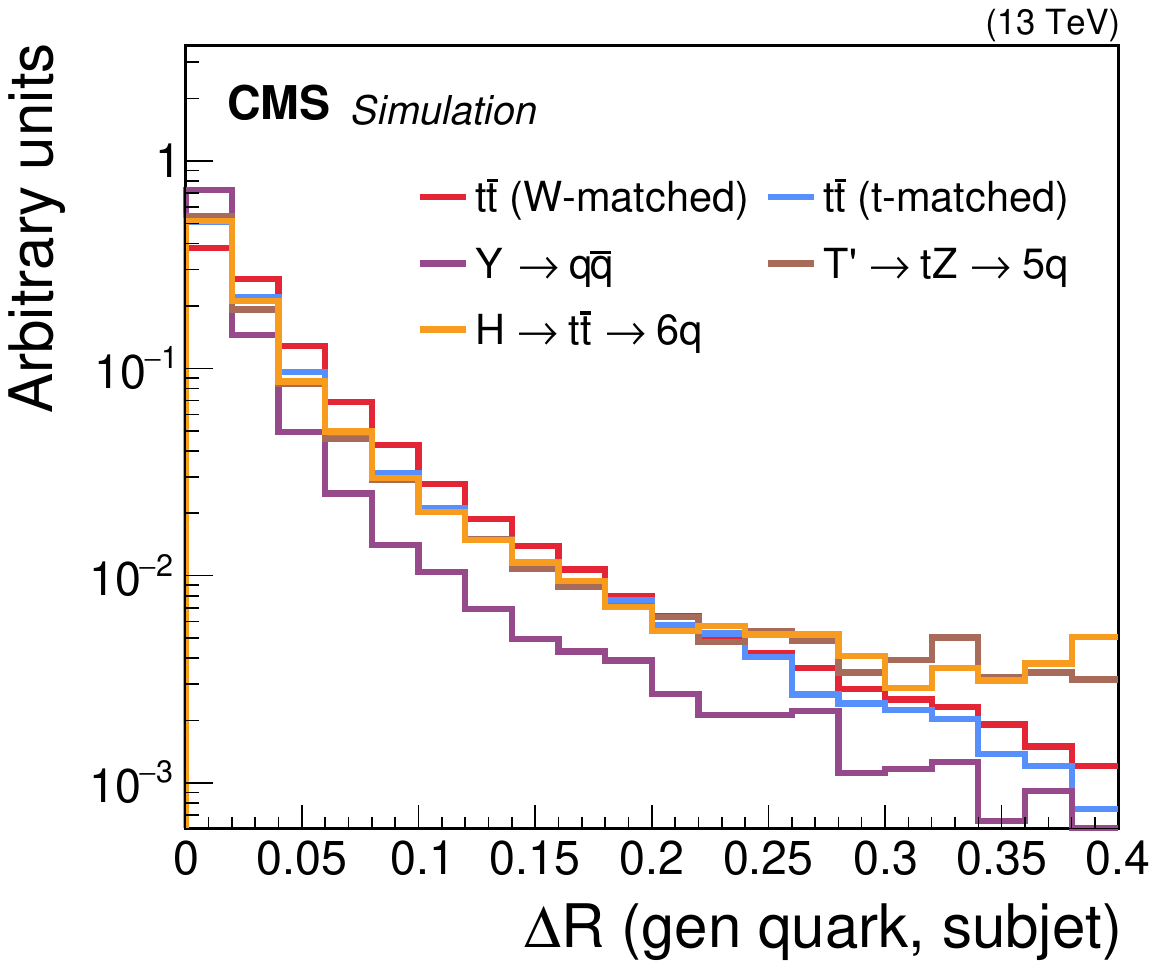}
  \caption{
    Distributions of the $\Delta R$ between subjets found by the reclustering procedure and closest generator-level quarks of the heavy resonance decay for various jet types. 
    The $\Delta R$ distributions for all signals peak towards zero, indicating that the reclustering procedure is performing well.
    }
  \label{fig:subjet_matching}
\end{figure}

Distributions of the $\Delta R$ between the subjets and generator-level quarks for jets of various types are shown in Fig. \ref{fig:subjet_matching}.
Overall, the reclustering performs quite well, even in these complicated jets with a high number of prongs, as evidenced by the distributions peaking towards zero.
The fraction of jets with one or more unclustered quarks is low in most signals, but grows with the number of prongs.
The fraction is $\approx$5\% and 10\% for \PW jets and top quark jets in the lepton+jets \ttbar region, respectively, $\approx$15\% for 4-pronged jets due to $\PR \to \PW\PW \to 4\PQq$, and $\approx$35\% for the 5- and 6-pronged jets due to $\PTpr \to \PQt \PZ \to 5 \PQq$ and $\PH \to \ttbar \to 6\PQq$, respectively.
The reclustering efficiency also has a nontrivial \pt dependence. 
Higher-\pt quarks are more efficiently reconstructed as subjets up to a point, but as the jet \pt increases, the prongs start to overlap and become more difficult to differentiate.

\subsection{Uncertainty due to the LJP distortion}

The LJP correction procedure assumes that the shower produced by an SM quark does not depend on whether it originates from a \PW jet or a different resonance decay with additional prongs. 
Such an assumption is necessary to apply the LJP density ratios derived from two-pronged $\PW$ jets to jets with higher prong multiplicities. 
However, this assumption has some limitations. 
The reclustering procedure may result in the assignment of a PF candidate to the wrong subjet, which would distort the splitting tree of the \kt algorithm and therefore the LJP density in a manner dependent on the number of prongs.
Alternatively, there could be color reconnection effects that differ between a given signal process and the $\PW$ jet sample, which could impact the emission structure and therefore the LJP density.
These effects can distort the LJP density of prongs from a given signal process versus the $\PW$ jet sample from which the data-simulation LJP density ratio is derived, and potentially alter the correction itself. 

To assess the uncertainty from LJP distortion, we compare the LJP densities derived from the original sample of $\PW$ jets, from which the nominal correction is derived, and from the simulated signal processes to which the correction is being applied.
The ratio of the two densities is used as the systematic uncertainty in the data-simulation LJP density ratio.

\subsection{Example uncertainty breakdown}

Table~\ref{tab:SF_uncs} shows the various uncertainties for jets from various processes. 
The uncertainties due to unclustered quarks and LJP distortion are generally dominant.
Despite the statistical uncertainties on the data-simulation LJP ratio being dominant, they have a reduced impact on the final correction as compared to systematic uncertainties because they are uncorrelated between different bins.
The correction to each subjet depends on ${\sim}O(10)$ independent bins. 
The impact of statistical uncertainties on the correction of an $N$-pronged jet are thus reduced by a factor of $\sim \sqrt{10 N}$, whereas the impact of systematic uncertainties is not. 
The \pt extrapolation uncertainty is generally larger for the BSM signal processes because the heavy resonance decays produce jets with large \pt. 
For the 2-prong $\PY \to \PQq\PQq$ and 6-prong $\PH \to \ttbar$ samples, the parent resonances have the same mass, but the former has a significantly larger \pt extrapolation uncertainty because the jet \pt is split among fewer subjets. 
The \PQb quark uncertainty is relatively small for all processes.

\begin{table}[!t]
  \renewcommand{\arraystretch}{1.3}
    \centering
    \topcaption{Uncertainties in the LJP reweighting scale factor for tagging jets from various processes. Uncertainties not applicable to a given process are denoted with a dash.
    }
    \cmsTable{
    \begin{tabular}{c c c c c c c}
      Jet type (\# prongs) & \PW (2) & \PQt (3) & $\PY \to \PQq\PQq $ (2) & $\PR \to \PW\PW$ (4) & $\PTpr \to \PQt \PZ$ (5) & $\PH \to \ttbar$ (6) \\
      \hline
    Selection variable & \scriptsize{\textsc{DeepAK8-MD} W} & $\tau_{32}$ & $\tau_{21}$ & \scriptsize{\textsc{DeepAK8-MD} WH} & $\tau_{43}$ & $\tau_{43}$ \\
      Ratio statistical & 0.01 & 0.01 & 0.01 & 0.02 & 0.01 & 0.02 \\
      Ratio systematic & 0.03 & 0.02 & 0.02 & 0.03 & 0.02 & 0.05 \\
      \pt extrapolation & \NA & 0.00 & 0.03 & 0.02 & 0.01 & 0.0 \\
      \PQb quark uncertainty & \NA & 0.01 & \NA & \NA & 0.01 & 0.01 \\
      Number of prongs & 0.01 & 0.03 & 0.00 & 0.02 & 0.03 & 0.03 \\
      Unclustered & $^{+0.02}_{-0.07}$ & $^{+0.06}_{-0.19}$ & $^{+0.01}_{-0.05}$ & $^{+0.07}_{-0.13}$ & $^{+0.05}_{-0.05}$ & $^{+0.10}_{-0.08}$ \\
      Distortion & \NA & $^{+0.11}_{-0.13}$ & $^{+0.11}_{-0.03}$ & $^{+0.13}_{-0.11}$ & $^{+0.11}_{-0.04}$ & $^{+0.14}_{-0.09}$ \\
      Total uncertainty & $^{+0.05}_{-0.07}$ & $^{+0.13}_{-0.22}$ & $^{+0.12}_{-0.07}$ & $^{+0.15}_{-0.14}$ & $^{+0.13}_{-0.07}$ & $^{+0.18}_{-0.13}$ \\
    \end{tabular}
    }
  \label{tab:SF_uncs}
\end{table}

\section{Tagging efficiency correction examples}
\label{sec:eff_examples}

\begin{figure}[!th]
  \centering
  \includegraphics[width = 0.7\textwidth]{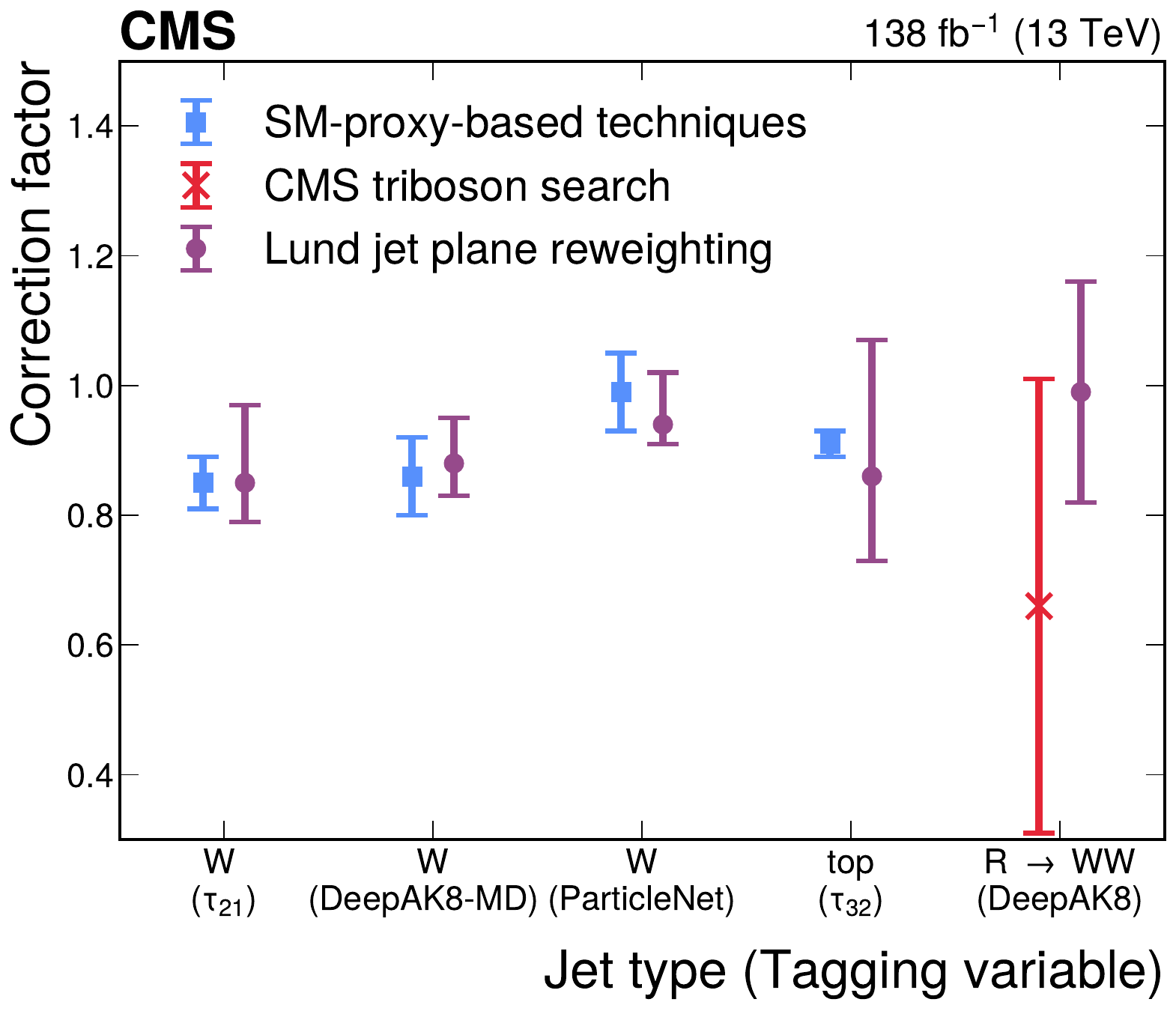}
  \caption{ 
    A comparison of correction factors for jet tagging efficiencies of various types, using standard calibration techniques based on SM proxy objects (blue squares), an extension of SM-proxy-based techniques using hard gluon radiation~\cite{triboson} (red crosses), and the LJP reweighting technique (purple squares). 
    The vertical error bars denote the uncertainty on each calibration technique. 
  }

\label{fig:SF_cmp}
\end{figure}

\begin{table}[!th]
  \centering
  \topcaption{
    A comparison of scale factors derived using the LJP correction procedure and other methods.  
    The scale factors derived with the LJP correction have larger uncertainties, but agree well with those from traditional methods.
    The comparison for the $\PR\to \PW\PW$ was taken from a recent search by the CMS Collaboration~\cite{triboson}.
  }
\begin{tabular}{c c c c c }
  Jet type (\# Prongs) & Selection & LJP & Comparison \\
  \hline
  \PW (2) &  $\tau_{21} < 0.34$ & $0.84 ^{+0.06}_{-0.12}$ & $0.85 \pm 0.04$  \\
  \PW (2) &  \textsc{DeepAK8-MD} ($\PW$) $>$ 0.7 & $0.88 ^{+0.05}_{-0.07}$ & $0.86 \pm 0.06$ \\
  \PW (2) &  \textsc{ParticleNet} ($\PW$) $>$ 0.94 & $0.94 ^{+0.03}_{-0.08}$ & $0.99 \pm 0.06$ \\
  \PQt (3) & $\tau_{32} < 0.52$ & $0.86 ^{+0.13}_{-0.21}$ & $0.91 \pm 0.02$ \\

  \hline

  $\PY \to \PQq\PQq $ (2) & $\tau_{21} < 0.34$ &  $0.81^{+0.13}_{-0.13}$  & \NA \\
  $\PR \to \PW\PW$ (4) & \textsc{DeepAK8-MD} WH $>$ 0.8 & $0.97^{+0.15}_{-0.14}$ & $0.66 \pm 0.35$ \\
  $\PTpr \to \PQt\PZ$ (5) & $\tau_{43} < 0.34$  & $0.82^{+0.13}_{-0.07}$ & \NA \\
  $\PH \to \PQt\PAQt$ (6) & $\tau_{43} < 0.34$ & $0.84^{+0.18}_{-0.13}$ & \NA \\
\end{tabular}
  \label{tab:SF_cmp}
\end{table}

Finally, we compare the performance of the LJP reweighting method for correcting the efficiency of substructure-based jet taggers versus prior methods based on measurements of SM proxy objects in data. 
For a given jet tagging criterion and simulated process, the corrected efficiency is obtained by including the LJP weights in the efficiency computation. 
Compared with the methods that use SM proxy objects in data, the LJP is more straightforward to use, since once the LJP density data-simulation ratios have been measured, the corrected efficiency is computed directly from the simulated samples. 
Further, by eliminating the need for an SM proxy object, the LJP method is applicable to jets with high-prong multiplicities and arbitrary phase space configurations.

To facilitate a comparison with other methods of calibrating the tagging efficiency, we compute a SF, defined as the ratio of the corrected efficiency to the uncorrected efficiency.
SFs for \PW jet and \PQt jet tagging are computed using the LJP reweighting correction procedure.
These SFs are compared with those obtained with standard procedures that directly measure the tagging efficiency in data, and are shown in Table~\ref{tab:SF_cmp} and Fig.~\ref{fig:SF_cmp}. 
Two-pronged jet tagging of \PW jets is performed using $\tau_{21}$, \textsc{DeepAK8-MD} score, and \textsc{ParticleNet-MD} score. 
Three-pronged jet tagging of top quark jets is performed using $\tau_{32}$. 
For four-pronged jets, the LJP method is compared with a proxy-jet method used in a search for $\PR\to\PW\PW\to4\PQq$~\cite{triboson}. 
Additionally, SFs are computed for BSM processes for which no SM proxy object has been observed in data. 
Two-pronged $\PY \to \PQq\PQq$ jets, selected using $\tau_{21}$, are produced with \pt larger than that available in the sample of \PW jets from lepton+jets \ttbar events. 
This process also highlights the \pt extrapolation procedure. 
An SF is also computed for six-pronged jets due to $\PH\to \PQt\PAQt$, selected using $\tau_{43}$, demonstrating the applicability to jets with very high prong-multiplicity. 

For the various \PW and \PQt tagging variables, the LJP SFs agree within uncertainties with the SFs obtained from direct measurements in data, generally with larger uncertainties.
For four-pronged tagging with the \textsc{DeepAK8-MD} tagger, the LJP SFs agree within the uncertainties with the proxy jet method, and has substantially smaller uncertainty.

\section{Summary}
\label{sec:summary}

A new method has been presented to improve the modeling in simulation of large-radius multiprong jets originating from the decay of heavy resonances into multiple quarks.
The method is based on a reclustering of the multiprong jet into separate subjets for each prong. 
The emissions of each subjet are corrected using the ratio of the Lund jet plane (LJP) densities between data and simulation, derived from a sample of \PW jets. 
The correction for the full jet is computed by combining the corrections of each of the subjets.
The method successfully improves the agreement between data and simulation of substructure observables of two-pronged \PW jets and three-pronged top quark jets. 
The LJP reweighting is also used to correct simulations using \PYTHIA for the parton shower to match \HERWIG, which validates that the correction performs well for jets with more than three prongs. 
The method can be used to correct the efficiency of substructure-based event selection criteria.
Efficiencies for \PW and \PQt tagging corrected with the LJP method agree well with the efficiencies measured directly in data. 
The main advance of the LJP method is that it can be applied to multiprong jets which could not be calibrated by previous methods.
It enables for the first time the calibration of jet tagging efficiencies for high-prong jets for which there are no comparable standard model processes of a high enough yield. 
The calibration of large-radius jets with high prong multiplicities enables the proper interpretation of the results of searches targeting such signatures.

\begin{acknowledgments}
 We congratulate our colleagues in the CERN accelerator departments for the excellent performance of the LHC and thank the technical and administrative staffs at CERN and at other CMS institutes for their contributions to the success of the CMS effort. In addition, we gratefully acknowledge the computing centers and personnel of the Worldwide LHC Computing Grid and other centers for delivering so effectively the computing infrastructure essential to our analyses. Finally, we acknowledge the enduring support for the construction and operation of the LHC, the CMS detector, and the supporting computing infrastructure provided by the following funding agencies: SC (Armenia), BMBWF and FWF (Austria); FNRS and FWO (Belgium); CNPq, CAPES, FAPERJ, FAPERGS, and FAPESP (Brazil); MES and BNSF (Bulgaria); CERN; CAS, MoST, and NSFC (China); MINCIENCIAS (Colombia); MSES and CSF (Croatia); RIF (Cyprus); SENESCYT (Ecuador); ERC PRG, RVTT3 and MoER TK202 (Estonia); Academy of Finland, MEC, and HIP (Finland); CEA and CNRS/IN2P3 (France); SRNSF (Georgia); BMBF, DFG, and HGF (Germany); GSRI (Greece); NKFIH (Hungary); DAE and DST (India); IPM (Iran); SFI (Ireland); INFN (Italy); MSIT and NRF (Republic of Korea); MES (Latvia); LMTLT (Lithuania); MOE and UM (Malaysia); BUAP, CINVESTAV, CONACYT, LNS, SEP, and UASLP-FAI (Mexico); MOS (Montenegro); MBIE (New Zealand); PAEC (Pakistan); MES and NSC (Poland); FCT (Portugal); MESTD (Serbia); MICIU/AEI and PCTI (Spain); MOSTR (Sri Lanka); Swiss Funding Agencies (Switzerland); MST (Taipei); MHESI and NSTDA (Thailand); TUBITAK and TENMAK (T\"{u}rkiye); NASU (Ukraine); STFC (United Kingdom); DOE and NSF (USA).

\hyphenation{Rachada-pisek} Individuals have received support from the Marie-Curie program and the European Research Council and Horizon 2020 Grant, contract Nos.\ 675440, 724704, 752730, 758316, 765710, 824093, 101115353, 101002207, 101001205, and COST Action CA16108 (European Union); the Leventis Foundation; the Alfred P.\ Sloan Foundation; the Alexander von Humboldt Foundation; the Science Committee, project no. 22rl-037 (Armenia); the Fonds pour la Formation \`a la Recherche dans l'Industrie et dans l'Agriculture (FRIA-Belgium); the Beijing Municipal Science \& Technology Commission, No. Z191100007219010, the Fundamental Research Funds for the Central Universities, the Ministry of Science and Technology of China under Grant No. 2023YFA1605804, and the Natural Science Foundation of China under Grant No. 12061141002 (China); the Ministry of Education, Youth and Sports (MEYS) of the Czech Republic; the Shota Rustaveli National Science Foundation, grant FR-22-985 (Georgia); the Deutsche Forschungsgemeinschaft (DFG), among others, under Germany's Excellence Strategy -- EXC 2121 ``Quantum Universe" -- 390833306, and under project number 400140256 - GRK2497; the Hellenic Foundation for Research and Innovation (HFRI), Project Number 2288 (Greece); the Hungarian Academy of Sciences, the New National Excellence Program - \'UNKP, the NKFIH research grants K 131991, K 133046, K 138136, K 143460, K 143477, K 146913, K 146914, K 147048, 2020-2.2.1-ED-2021-00181, TKP2021-NKTA-64, and 2021-4.1.2-NEMZ\_KI-2024-00036 (Hungary); the Council of Science and Industrial Research, India; ICSC -- National Research Center for High Performance Computing, Big Data and Quantum Computing, FAIR -- Future Artificial Intelligence Research, and CUP I53D23001070006 (Mission 4 Component 1), funded by the NextGenerationEU program (Italy); the Latvian Council of Science; the Ministry of Education and Science, project no. 2022/WK/14, and the National Science Center, contracts Opus 2021/41/B/ST2/01369 and 2021/43/B/ST2/01552 (Poland); the Funda\c{c}\~ao para a Ci\^encia e a Tecnologia, grant CEECIND/01334/2018 (Portugal); the National Priorities Research Program by Qatar National Research Fund; MICIU/AEI/10.13039/501100011033, ERDF/EU, "European Union NextGenerationEU/PRTR", and Programa Severo Ochoa del Principado de Asturias (Spain); the Chulalongkorn Academic into Its 2nd Century Project Advancement Project, and the National Science, Research and Innovation Fund via the Program Management Unit for Human Resources \& Institutional Development, Research and Innovation, grant B39G680009 (Thailand); the Kavli Foundation; the Nvidia Corporation; the SuperMicro Corporation; the Welch Foundation, contract C-1845; and the Weston Havens Foundation (USA).
\end{acknowledgments}

\bibliography{auto_generated} 
\cleardoublepage \appendix\section{The CMS Collaboration \label{app:collab}}\begin{sloppypar}\hyphenpenalty=5000\widowpenalty=500\clubpenalty=5000
\cmsinstitute{Yerevan Physics Institute, Yerevan, Armenia}
{\tolerance=6000
A.~Hayrapetyan, V.~Makarenko\cmsorcid{0000-0002-8406-8605}, A.~Tumasyan\cmsAuthorMark{1}\cmsorcid{0009-0000-0684-6742}
\par}
\cmsinstitute{Institut f\"{u}r Hochenergiephysik, Vienna, Austria}
{\tolerance=6000
W.~Adam\cmsorcid{0000-0001-9099-4341}, J.W.~Andrejkovic, L.~Benato\cmsorcid{0000-0001-5135-7489}, T.~Bergauer\cmsorcid{0000-0002-5786-0293}, K.~Damanakis\cmsorcid{0000-0001-5389-2872}, M.~Dragicevic\cmsorcid{0000-0003-1967-6783}, C.~Giordano, P.S.~Hussain\cmsorcid{0000-0002-4825-5278}, M.~Jeitler\cmsAuthorMark{2}\cmsorcid{0000-0002-5141-9560}, N.~Krammer\cmsorcid{0000-0002-0548-0985}, A.~Li\cmsorcid{0000-0002-4547-116X}, D.~Liko\cmsorcid{0000-0002-3380-473X}, I.~Mikulec\cmsorcid{0000-0003-0385-2746}, J.~Schieck\cmsAuthorMark{2}\cmsorcid{0000-0002-1058-8093}, R.~Sch\"{o}fbeck\cmsAuthorMark{2}\cmsorcid{0000-0002-2332-8784}, D.~Schwarz\cmsorcid{0000-0002-3821-7331}, M.~Shooshtari, M.~Sonawane\cmsorcid{0000-0003-0510-7010}, W.~Waltenberger\cmsorcid{0000-0002-6215-7228}, C.-E.~Wulz\cmsAuthorMark{2}\cmsorcid{0000-0001-9226-5812}
% ...
\par}
\vskip\cmsinstskip
\dag:~Deceased\\
$^{1}$Also at Yerevan State University, Yerevan, Armenia\\
$^{2}$Also at TU Wien, Vienna, Austria\\
% ...
\end{sloppypar}
\end{document}